\documentclass[letterpaper]{JHEP3}

\usepackage{amssymb}
\usepackage{graphicx}

\newcommand{\bea}{\begin{eqnarray}}
\newcommand{\eea}{\end{eqnarray}}
\newcommand{\beq}{\begin{equation}}
\newcommand{\eeq}{\end{equation}}
\newcommand{\nn}{\nonumber}
\newcommand{\ignore}[1]{}

\def\q{q'}
\def\r{r'}
\def\s{s'}

\def\bfk{{\bf k}}
\def\bfx{{\bf x}}
\def\bfy{{\bf y}}

\def\nr{\varrho}

\def\ba{\overline{a}}

\def\bc{\overline{\chi}}
\def\be{\overline{\eta}_c}

\def\bU{\overline{\Upsilon}}

\def\bbfx{\overline{\bf x}}

\def\cF{{\cal F}}
\def\cN{{\cal N}}

\def\bcF{\overline{\cal F}}
\def\bC{\overline{C}}
\def\btF{\tilde{\cal F}}

\def\<{\langle}
\def\>{\rangle}

\title{Observable effects of anisotropic bubble nucleation}

\author{Jose J.~Blanco-Pillado and Michael P.~Salem\\
Institute of Cosmology, Department of Physics and Astronomy,  
Tufts University, Medford, Massachusetts 02155, USA}

\abstract{Our universe may have formed via bubble nucleation in an 
eternally-inflating background.  Furthermore, the background may have 
a compact dimension---the modulus of which tunnels out of a metastable 
minimum during bubble nucleation---which subsequently grows to become 
one of our three large spatial dimensions.  Then the reduced symmetry 
of the background is equivalent to anisotropic initial conditions in 
our bubble universe.  We compute the inflationary spectrum in such a 
scenario and, as a first step toward understanding the effects of 
anisotropy, project it onto spherical harmonics.  The resulting 
spectrum exhibits anomalous multipole correlations, their relative 
amplitude set by the present curvature parameter, which appear to 
extend to arbitrarily large multipole moments.  This raises the 
possibility of future detection, if slow-roll inflation does not last 
too long within our bubble.  A full understanding of the observational 
signal must account for the effects of background anisotropy on photon 
free streaming, and is left to future work.}

\begin{document}

\section{Introduction} 
\label{sec:introduction}

Inflation is generically eternal \cite{V83,L86}.  That is, for 
many scalar field potentials the physical volume of inflating 
spacetime is divergent, with inflation ending only in localized 
``pockets'' within the inflating background.  This is the case, for 
instance, when inflation is driven by the positive vacuum energy of 
some metastable ``parent'' vacuum, in which the vacuum phase of our 
``daughter'' universe arises due to tunneling through a potential 
barrier.  The tunneling process is described by an instanton that 
interpolates between parent and daughter vacua, and appears as 
bubble nucleation in the inflating spacetime \cite{CDL,LW}.  Indeed
this view of cosmology is supported by the present understanding of 
string theory, which argues for the existence of an enormous 
landscape of such metastable vacua \cite{BP,S03,D,DK}.     

In the standard picture, both the parent and daughter vacua have 
three large (expanding) spatial dimensions.  Then the symmetries of 
the de Sitter parent vacuum suggest the daughter bubble should 
possess a homogeneous and isotropic geometry---in particular it 
should possess an O(3) rotational symmetry on any homogeneous 
foliation \cite{CDL}.  Yet the number of large spatial dimensions 
may vary from vacuum to vacuum \cite{G,B-PS-PV,CJR,B-PS-PV-2}, as is 
expected in string theory.  In particular, one of the large spatial 
dimensions of the daughter vacuum may be compact in the parent 
vacuum.  The size of this dimension can be characterized 
by a ``volume'' modulus that, during bubble nucleation, tunnels out 
of a metastable minimum, and subsequently grows to very large values.  
While the resulting bubble is still assumed to reflect the 
symmetries of the parent vacuum, the presence of the compact 
dimension breaks O(3) rotational invariance.  Indeed, we expect the 
bubble geometry to be toroidal, with O(2) rotational symmetry in the 
two large spatial dimensions, uniformly wrapping around the compact 
space.  This would appear as anisotropic initial conditions to an 
observer in the daughter vacuum.   

While the formerly-compact spatial dimension remains globally closed 
in the daughter bubble, this will not necessarily be evident to a 
local observer.  In order for the daughter vacuum to 
correspond to our universe, its local evolution should approach 
an approximately O(3) rotationally-symmetric Friedmann-Robinson-Walker 
(FRW) cosmology, and the circumference of each large spatial dimension 
should become and/or remain much larger than the Hubble radius.  
In fact these conditions are easy to satisfy---the latter is akin to 
solving the horizon problem of classical big bang cosmology, and is 
accomplished by a sufficiently long period of slow-roll inflation 
(after bubble nucleation) along each large spatial dimension 
\cite{Guth81}.  (A period of inflation after bubble nucleation is 
required even in the standard O(3)-symmetric situation, in order to 
redshift away the large initial spatial curvature of the bubble.)  
Meanwhile, it is well known that during such inflation an initially 
homogeneous but anisotropic universe rapidly approaches local
isotropy \cite{BT,W}.   

Although the initial anisotropy rapidly redshifts away, background 
anisotropy present at the onset of inflation will generate statistical 
anisotropy in quantum fluctuations as they expand beyond the Hubble 
radius, and this in turn can modify the spectrum of primordial density 
perturbations \cite{GCP,ACW,PPU,WKS10}.  During inflation the affected 
modes are pushed to physical scales far beyond the local horizon; 
however if the duration of inflation is appropriate they will have 
re-entered in time to form the largest observable scales in the cosmic 
microwave background (CMB).  Indeed, seemingly anomalous correlations 
have already been detected among the low-multipole CMB anisotropies 
\cite{O-CTZH,SSHC,EHBGL,LM,JBEGH}, which might indicate deviations 
from statistical isotropy in the inflationary spectrum. (It should be 
noted that the significance of these ``anomalies'' is difficult 
to assess, and their source(s) could be non-cosmological.)  In this 
context a number of models of anisotropy during inflation have been 
proposed; see e.g.~\cite{BBK05,DDR,A-P,BM08,DGW,GMV,YS08,WKS09,DM10}.  
However unlike other approaches, anisotropic bubble nucleation provides 
concrete theoretical constraints on the form of initial conditions, 
and serves as a natural extension of the standard inflationary 
scenario.  Furthermore, it provides an opportunity to confirm aspects 
of the landscape/multiverse hypothesis.  

When the background is homogeneous and isotropic, inflationary 
perturbations generally decouple into scalar, vector, and tensor 
modes (for a review of cosmological perturbation theory see 
e.g.~\cite{MFB}).  The same cannot be said for fluctuations about 
anisotropic backgrounds, which complicates the corresponding analysis.  
For this reason we assume that metric perturbations are suppressed, 
until the background geometry of the bubble has become essentially 
isotropic, after which standard cosmological perturbation theory can 
be used.  As a concrete model one can imagine the spectrum of 
isocurvature fluctuations in a sub-dominant scalar field, which are
much later converted into adiabatic density perturbations (as in the
curvaton mechanism \cite{curvaton}).  Note that this simplification 
comes at little cost:  we are interested not in the (model-dependent) 
amplitude or tilt of the spectrum, but in its statistical anisotropy, 
and this should not depend strongly on the back-reaction of the 
scalar on the metric.

Even with this and some other simplifications, the analysis is 
rather complicated.  Although we obtain an analytic expression for 
the inflationary power spectrum in terms of an appropriate set of 
anisotropic mode functions, we must resort to numerical evaluation 
to project this spectrum onto spherical harmonics.  Still, we find 
that certain patterns are evident:  whereas the standard (isotropic)
picture gives a multipole correlator 
$C_{\ell\ell'mm'}=\<\hat{a}_{\ell m}^{\phantom{\dagger}}\,
\hat{a}^\dagger_{\ell'm'}\>$ that is diagonal in both $\ell$ and 
$\ell'$ and in $m$ and $m'$, and is independent of $m$, our scenario 
introduces off-diagonal components in $\ell$ and $\ell'$ (when 
$\ell-\ell'=\pm 2$), and introduces $m$-dependence into 
$C_{\ell\ell'mm'}$ (it is still diagonal in $m$ and $m'$).  (These 
results are not unlike those of \cite{ACW}, which studied a Bianchi 
type I anisotropic cosmology.)  Our approximations limit us to a 
region in parameter space where the corrections to $C_{\ell\ell'mm'}$ 
are suppressed relative to the leading order terms by roughly the 
present-day curvature parameter $\Omega^0_{\rm curv}$.  While this 
greatly constrains the size of these effects, they appear to extend to 
arbitrarily large $\ell$, giving hope for statistically significant 
future detection \cite{PK}.  Note that because the statistically 
anisotropic contributions to $C_{\ell\ell'mm'}$ are suppressed by 
$\Omega^0_{\rm curv}$.

We have focused on statistical anisotropies in inflationary 
perturbations, however in this scenario the background spatial 
curvature of the bubble is itself anisotropic.  In particular, the 
spatial geometry is flat along one direction and open in the 
two-dimensional planes orthogonal to that direction.  While our 
computation of the inflationary spectrum accounts for this spatial 
curvature anisotropy, our projection onto spherical harmonics does 
not.  Indeed, anisotropic spatial curvature induces anisotropic 
expansion, which affects the free streaming of photons and thus 
deforms the surface of last scattering, along with our perception of 
angular scales on it \cite{DD}.  This affects the appearance of the 
inflationary spectrum, inducing corrections to the observed multipole
correlator $C_{\ell\ell'mm'}$.  A full understanding of the 
observable signatures of this model involves combining both of these 
effects; this is left to future work \cite{B-PS2}.

The remainder of this paper is organized as follows (an effort has
been made to make the major sections self-contained).  We study the 
dynamics of anisotropic bubble nucleation, within the context of a 
toy model of modulus stabilization, in Section \ref{sec:bubble}.  
The primary goal of this section is to obtain the instanton boundary 
conditions that determine the initial conditions for the subsequent
bubble evolution.  However, because our compactification of one 
extra dimension with positive vacuum energy is (to our knowledge) 
novel, we present the model in some detail.  In Section 
\ref{sec:background} we describe the salient features of the 
post-nucleation, background evolution of the bubble, focusing on the
(pre-)inflationary geometry (including obtaining a simple analytic 
approximation of the metric).  In Section \ref{sec:perts} 
we compute the spectrum of inflationary perturbations in a massless 
scalar field.  To better understand the observational signatures of
this spectrum, we here also perform a basic analysis of its 
projection onto spherical harmonics.  Some issues pertaining to the 
plausibility of observing this scenario are discussed in 
Section~\ref{sec:likelihood}.  Meanwhile a final summary, including 
a discussion of various avenues for future work, is provided in 
Section~\ref{sec:discussion}.    

\vspace{11pt}
\noindent
Preliminary accounts of this work were presented in \cite{MPS}.  

\section{Anisotropic bubble nucleation}
\label{sec:bubble}

We study the possibility that our universe formed via bubble 
nucleation within some parent vacuum in which one of the large 
spatial dimensions of our vacuum is 
compact.\footnote{\label{foot:1}One might also consider 
the possibility that two of the large spatial dimensions of our 
vacuum are compactified in the parent vacuum.  In this case the 
volume modulus couples to the Ricci scalar, but because the 
(1+1)-dimensional gravity of the non-compact dimensions is 
conformally invariant, the theory cannot be transformed to the 
Einstein frame.  Although this by itself poses no formal problems 
to constructing viable models, it does complicate the analysis, 
and so to retain focus we leave this possibility to future work.}  
Although string theory indicates that our vacuum itself has six 
or seven compact dimensions, for simplicity we consider the 
associated moduli fields to be non-dynamical spectators in all of 
the processes of interest here.  Thus the parent vacuum is taken 
to have one compact dimension, the ``volume'' modulus of which 
tunnels out of a metastable minimum during bubble nucleation, 
which subsequently grows to very large size to create our 
effectively (3+1)-dimensional daughter vacuum.  Note that the 
metastable minimum referred to above must have positive vacuum 
energy, so the parent vacuum can decay to our  
positive-vacuum-energy universe.

\subsection{(2+1+1)-dimensional modulus stabilization}
\label{ssec:compactification}

We first construct an explicit (toy) model of compactification.  
The model is based on one in \cite{B-PS-PV} (see also references
therein).  Our purpose is to demonstrate that there are no basic 
dynamical obstacles to implementing our picture of bubble nucleation, 
and meanwhile to provide a concrete model for future reference.  In 
order to generate a metastable, positive-vacuum-energy solution 
for the volume modulus, we use the winding number of a complex 
scalar field to stabilize the size of the compact dimension.  In 
particular, we consider the (3+1)-dimensional (hereafter denoted 
4d) action
\beq
S = \int\! \sqrt{-g}\,d^4x 
\left[ \frac{1}{16\pi G}\left( R - 2\Lambda\right) 
-\frac{1}{2}K(\partial_\mu\varphi^*\partial^\mu\varphi) -  
\frac{\lambda}{4}\left(|\varphi |^2-\eta^2\right)^2\right] ,
\label{4daction}
\eeq  
where $g$ is the determinant of the 4d metric $g_{\mu\nu}$, $R$ 
is the 4d Ricci scalar, $\Lambda$ is a cosmological constant, and 
$\varphi$ is a complex scalar field for which we allow a 
non-canonical ``kinetic'' function specified by $K$.  The other 
terms are constants.  Other degrees of freedom, for instance the 
inflaton and the matter fields of the Standard Model, are assumed 
to be unimportant during the tunneling process, and are absorbed 
into $\Lambda$ (and/or $g$ and $R$).

The stabilization of the volume modulus of a compact dimension 
$z$ is most conveniently studied using a metric ansatz with line 
element
\beq
ds^2 = e^{-\Psi}\,\overline{g}_{ab}\,dx^adx^b 
+ L^2e^{\Psi}\,dz^2 \,,
\label{ansatz}
\eeq   
where $\Psi$ represents the modulus field.  The effective 
(2+1)-dimensional (hereafter 3d) metric $\overline{g}_{ab}$ and the 
modulus $\Psi$ are both taken to be independent of $z$.  Meanwhile, 
the compact dimension $z$ is defined using periodic boundary conditions, 
with $0\leq z\leq 2\pi$, so that it has the topology of a circle with 
physical circumference $2\pi L\,e^{\Psi/2}$.  Note that we have 
introduced the following notation.  Any quantity defined explicitly 
within the effective 3d theory (i.e.~the theory with the $z$ dimension 
integrated out), such as the effective 3d metric, is marked with an 
overline.  Whereas Greek indices are understood to run over all 
dimensions, Latin indices are understood to run over all but the 
$z$ dimension.

We seek a solution for the scalar field $\varphi$ that stabilizes 
the modulus $\Psi$ with respect to small perturbations.  The 
action (\ref{4daction}) gives the equation of motion
\beq
\partial_\mu\left[\sqrt{-g}\,K'\,\partial^\mu\varphi \right]
- \sqrt{-g}\,\lambda\varphi\left(|\varphi|^2-\eta^2\right) = 0\,,
\label{phieom}
\eeq
where $K'\equiv dK(X)/dX$, with 
$X=\partial_\mu\varphi^*\partial^\mu\varphi$.  Consider for the 
moment that $\Psi$ is a constant, $\Psi=\Psi_p$.  Then the above 
equation of motion permits the solution 
\beq
\varphi=\left(\eta^2 -\frac{n^2}{\lambda L^2}K'
e^{-\Psi_p}\right)^{\!\!1/2}\! e^{inz} \,,
\label{phisol}
\eeq
where $n$ is an integer, representing the winding number of the 
phase of $\varphi$.  Ultimately we are interested in the dynamics 
of $\Psi$, in which case $\Psi$ depends on time and (\ref{phisol}) 
is no longer an exact solution to the equation of motion.  We 
brush this complication aside by taking
\beq
\eta^2 \gg \frac{n^2}{\lambda L^2}K'e^{-\Psi} \,, 
\label{etaapprox}
\eeq
so that to leading order we have $\varphi = \eta\,e^{inz}$, which 
in turn gives
\beq
X\equiv\partial_\mu\varphi^*\partial^\mu\varphi
=\frac{n^2\eta^2}{L^2}e^{-\Psi} \,.
\eeq
Note that we are only interested in dynamics that increase the 
size of the compact dimension, i.e.~dynamics that increase 
$\Psi$.  Therefore if the above inequality is valid in the parent 
vacuum, it is valid throughout our analysis (we assume that $K'(X)$ 
contains no poles in $X$).  The effective 3d action is then 
obtained by using this solution to integrate the $z$ dimension out 
of the action.  After integrating by parts, this gives
\beq
S_{\rm 3d} = \int\! \sqrt{-\overline{g}}\,d^3x\left[ 
\frac{1}{16\pi \overline{G}}\,\overline{R}
-\frac{1}{2}\partial_a\psi\partial^a\psi 
-\frac{\Lambda}{8\pi \overline{G}}\,e^{-\alpha\psi}
-\frac{1}{2}e^{-\alpha\psi} \overline{K}(X(\psi))\right]\,,
\label{3daction}
\eeq   
where we have defined $\overline{G}\equiv G/(2\pi L)$, 
$\overline{K}\equiv 2\pi L\, K$, and $\psi\equiv\Psi/\alpha$, where 
$\alpha=\sqrt{16\pi \overline{G}}$, with the rescaled modulus $\psi$ 
being a ``canonical'' scalar field in the 3d theory.  

Recall that our goal is to find a theory/solution in which the 
modulus $\psi$ is stabilized with positive vacuum energy.  The 
satisfaction of these conditions can be verified by studying the 
effective potential of $\psi$, which corresponds to the last two
terms of (\ref{3daction}).  It is sufficient for our purposes to 
identify a single viable model, so let us propose
\beq
\overline{K}(X) = 2 \pi L\left( X + \kappa_2 X^2 + \kappa_3 X^3\right)\,,
\label{Kdef}
\eeq  
where $\kappa_2$ and $\kappa_3$ are constants.  Note that it is 
inappropriate to consider this form of kinetic function as the 
truncation of a longer series, since we will rely on the various
terms being comparable in size.  In this sense, (\ref{Kdef}) is 
ad hoc.  The choice $K(X)=2\beta\sqrt{1+\gamma X}-2\beta$, with 
$\beta$ and $\gamma$ being constants, may seem more natural, 
since kinetic terms of this form arise in certain braneworld 
scenarios.  However we have found that this kinetic term does not 
permit any metastable solutions.  Nevertheless, we do not consider 
this a serious issue.  We are here focusing on only one very simple 
model of compactification, as a representative of the essential 
dynamics.  It seems reasonable to expect that a thorough 
investigation of, say, the string landscape, would reveal a large 
number of realistic possibilities.

The 3d effective potential arising due to our choice of 
$\overline{K}(X)$ is
\beq
\overline{V}(\psi) = \frac{\Lambda}{8\pi \overline{G}}\,e^{-\alpha\psi}
+ 2 \pi L \left(\frac{n^2\eta^2}{2L^2} \,e^{-2\alpha\psi}
+\frac{\kappa_2\,n^4\eta^4}{2L^4}\,e^{-3\alpha\psi}
+\frac{\kappa_3\,n^6\eta^6}{2L^6}\,e^{-4\alpha\psi}\right)  \,.
\label{Veff}
\eeq
Let us now explain the choice of $\overline{K}$ leading to the above 
result.  If we had used only the canonical term, $K(X)=X$, then the 
effective potential would have had no stationary points at 
finite $\psi$ (recall that we require $\Lambda>0$).  Including 
the term proportional to $\kappa_2$ creates a stationary point 
for some negative values of $\kappa_2$, but it is always a 
local maximum.  Thus it takes two terms in addition to the 
canonical one to allow for a local minimum.  

\begin{figure}[t!]
\begin{center}
\includegraphics[width=0.5\textwidth]{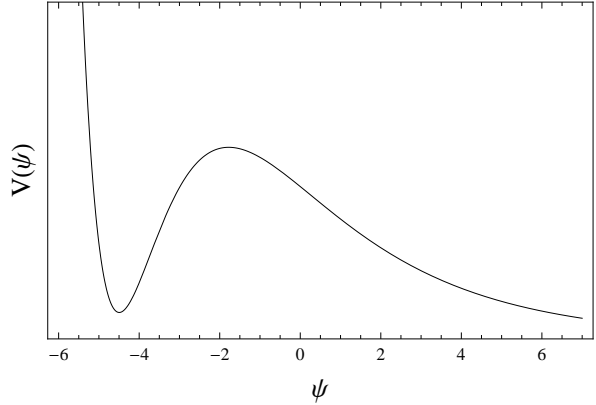} 
\caption{\label{fig:V}The effective potential of the modulus 
field $\psi$; see text for details.}
\end{center}
\end{figure}

The potential $\overline{V}(\psi)$ of (\ref{Veff}) is displayed in 
Figure~\ref{fig:V} for a particular set of parameter values:  
$\Lambda/8\pi= 10^{-11}$, $X_0=n^2\eta^2/L^2=10^{-11}$, 
$\kappa_2=-X_0^{-1}$, and $\kappa_3=0.16X_0^{-2}$ (all quantities 
are given in units of $G$).  The value of $\Lambda$ is chosen to 
roughly correspond to the upper-bound observational limit of the 
inflationary potential energy in our universe \cite{WMAP}, while 
$X_0$ is simply chosen to be on the same order as $\Lambda$.  The 
other parameters, namely the coefficients that determine $\kappa_2$ 
and $\kappa_3$, were simply guessed by trial and error.  Since 
$n$, $\eta$, and $L$ appear only in the combination 
$X_0=n^2\eta^2/L^2$, there is no technical problem with 
satisfying the inequality (\ref{etaapprox}) for any shape of 
potential $\overline{V}(\psi)$.  Also, we have checked that the 
above values of $\kappa_2$ and $\kappa_3$ satisfy the ``hyperbolic 
condition'' of \cite{EB06}, indicating that the underlying 
solution for $\varphi$ is stable to small perturbations.  
     
The effective potential displayed in Figure~\ref{fig:V} has a 
positive vacuum energy local minimum at some value $\psi=\psi_p$.  
The state with $\psi=\psi_p$ is the parent vacuum, which appears 
as effectively 3d de Sitter space on scales much larger than 
$2\pi L\,e^{\alpha\psi_p/2}$.  The daughter vacuum is created when 
$\psi$ tunnels through the barrier, to some value $\psi=\psi_d$, 
after which $\psi$ accelerates from rest and rolls down the 
potential, with $\psi\to\infty$ as time $x^0\to\infty$.  

It is important to this work that the daughter vacuum be created 
via tunneling (Coleman--De Luccia instanton), as opposed to by 
quantum diffusion over the barrier (Hawking-Moss instanton 
\cite{HawkingMoss}), because the former gives us a handle on the 
initial geometry of the daughter universe.  This translates to 
the requirement that $|V''| \gtrsim 32\pi G\, V$ at the peak of 
the potential barrier.  This condition not hard to satisfy; in 
particular it is satisfied for the choice of parameters given 
above.  Note that this is not the only channel by which the parent 
vacuum may decay---in particular it may transition to any vacuum 
with winding number $n'\ne n$ (see e.g.~\cite{B-PS-PV}).  
Insofar as $\psi$ remains trapped in a local minimum, these 
transitions are of no interest, as observers like us could never 
arise in the resulting 3d daughter vacuum.  If such a transition 
eliminates the local minimum or otherwise sets $\psi$ free, then 
we simply consider it a more sophisticated version of the case we 
study here.

\subsection{Tunneling instanton and bubble geometry}
\label{ssec:instanton}

The semi-classical theory of vacuum decay via bubble nucleation 
is laid out in \cite{CDL}, and we find that we can work in direct 
analogy to that analysis.  The tunneling instanton is 
found by studying the Euclidean action, and we begin by 
constructing a metric ansatz that exploits the full symmetry of 
the parent vacuum.  In the 3d effective theory with the compact 
dimension integrated out, this gives the Euclidean line element
\beq
ds^2=d\xi^2
+\rho^2(\xi)\left[d\chi^2+\sin^2(\chi)\,d\phi^2 \right] \,, 
\label{Emetric}
\eeq  
where $\xi$ is a radial coordinate and $\chi$ and $\phi$ are 
angular coordinates on a two-sphere of radius $\rho(\xi)$.  An
effective theory constructed by integrating out a compact 
dimension is generally valid only on scales much larger than 
that of the compact dimension.  In the present case, however,
the tunneling field is also the modulus of the $z$ dimension.  
If the 4d manifold factorizes as in our ansatz (\ref{ansatz}), 
then the modulus $\psi$ is independent of $z$ on all scales.
Therefore any domain wall of $\psi$ must be independent of $z$, 
and there can be no topological structures to support any 
$z$ dependence of $\rho$ in the tunneling instanton.  (These 
statements hold in particular because the tunneling field is 
the modulus of the compact dimension; of course such 
restrictions would not apply to a generic instanton in this 
background.)  

After integrating out the coordinates $\chi$ and $\phi$ (and 
$z$), the Euclidean action can be written
\beq
S_{\rm E} = 4\pi\!\int d\xi \left[ -\frac{1}{8\pi {\overline G}}
\left( \dot{\rho}^2+1\right) + \rho^2 \left(\frac{1}{2}\dot{\psi}^2 
+ \overline{V}(\psi)\right)\right] ,
\label{Eaction}
\eeq
where, in the present context, the dot denotes differentiation 
with respect to $\xi$.  The field equations of the above action 
can be written
\bea
\frac{\dot{\rho}^2}{\rho^2} -\frac{1}{\rho^2} &=& 8\pi \overline{G}
\left(\frac{1}{2}\dot{\psi}^2 - \overline{V}\right) \label{E}\\
\ddot{\psi} + 2\frac{\dot{\rho}}{\rho}\dot{\psi} 
&=& \overline{V}'(\psi) \,,\phantom{\bigg(\bigg)} \label{F}
\eea
where the prime denotes differentiation with respect to $\psi$.
These are the same as the field equations of the standard 4d 
tunneling instanton, except for some numerical factors \cite{CDL}.  
In the present case, the instanton interpolates between a local 
excitation of the parent vacuum, in which $\psi=\psi_p$, and some 
point on the opposite side of the potential barrier, at which 
$\psi=\psi_d$.  Inside the daughter vacuum $\psi$ continues to 
grow, but this evolution is not covered by the tunneling 
instanton---the geometry of this region is deduced by matching at 
the instanton boundary, where $\psi=\psi_d$.  As a consequence, 
the circumference of the compact dimension never exceeds 
$2\pi L\,e^{\alpha\psi_d}$ in the region covered by the instanton.

\begin{figure}[t!]
\begin{center}
\begin{tabular}{cc}
\includegraphics[width=0.475\textwidth]{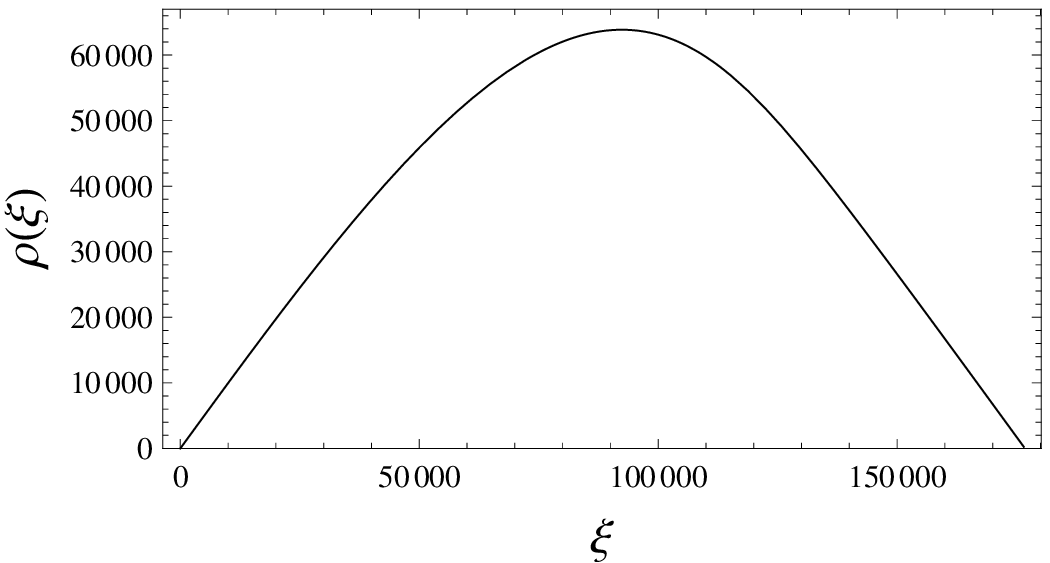} &
\includegraphics[width=0.457\textwidth]{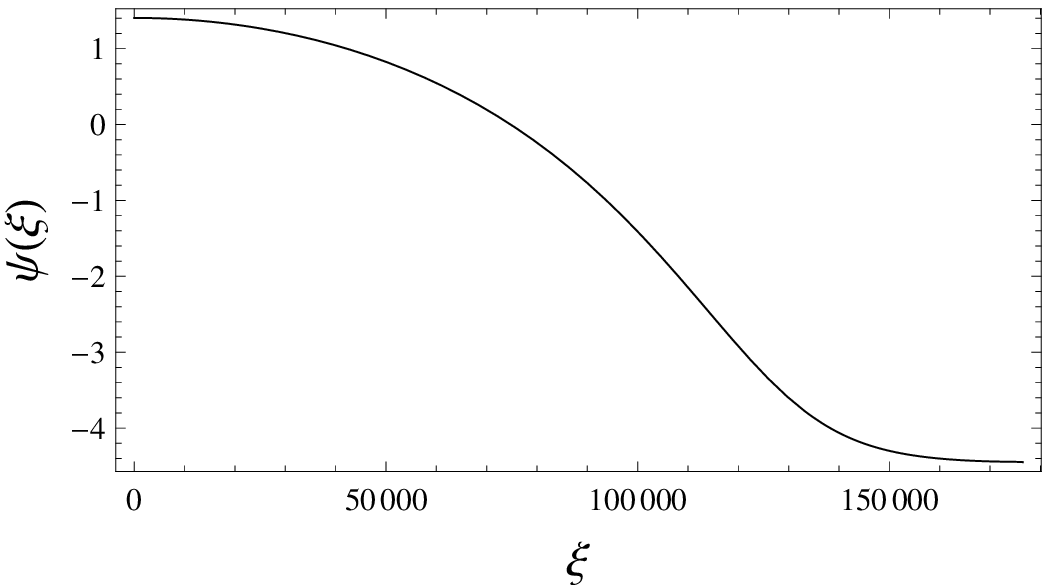} \\
\end{tabular}
\caption{\label{fig:instanton3d}The 3d instanton solution 
$\rho(\xi)$ (left panel) and $\psi(\xi)$ (right panel), all quantities
given in units of $G$.  The large numbers are due to the dynamics being 
sub-Planckian; see text for details.}
\end{center}
\end{figure}

It is straightforward to verify the existence of a tunneling
instanton by direct numerical integration.  The boundary conditions
are determined by Taylor expanding $\rho(\xi)$ and $\psi(\xi)$, 
\bea
\rho(\xi) &=& \xi - \frac{4\pi\overline{G}}{3}\,\overline{V}(\psi_d)\,
\xi^3 + \ldots \label{boundary1}\\
\psi(\xi) &=& \psi_d - 
\frac{\pi\overline{G}}{3}\,\psi_d\,\overline{V}'(\psi_d)\,
\xi^2 + \ldots \,, \label{boundary2}
\eea
where the coefficients of the expansions are determined using the 
equations of motion.  The value of $\psi_d$ is set by trial and error, 
so that as $\xi$ approaches some value $\xi_{\rm max}$, the instanton 
smoothly approaches $\psi(\xi)\to$ constant (the excited value just 
prior to tunneling), and likewise $\rho(\xi)\to\xi_{\rm max}-\xi$.  Of 
course the details of the instanton depend on the shape of the modulus 
potential $\overline{V}$; to demonstrate the consistency of our model 
we use the same values of parameters as are used to generate Figure 
\ref{fig:V}.  The results of this numerical integration are displayed 
in Figure \ref{fig:instanton3d}, and agree with the qualitative 
description above (note that $\xi=0$ corresponds to the side of the 
potential barrier leading into the daughter vacuum).  In Appendix 
\ref{sec:instanton} we analyze this tunneling solution from the 4d 
perspective, substantiating these results.

The instanton displayed in Figure \ref{fig:instanton3d} does not 
appear to be well-described by the so-called thin-wall approximation.  
Nevertheless, it is not implausible that other modulus potentials 
exist in which the thin-wall approximation is accurate.  In such case
the analysis can proceed in exact analogy to that of \cite{CDL}.  
Although we do not here propose such a model, we shall assume the 
tunneling instanton can be well-described by the thin-wall approximation
to simplify the computation of inflationary perturbations in Section 
\ref{sec:perts} (Appendix \ref{sec:BD}).

Finally, let us note the geometry of the nucleating bubble \cite{CDL}.  
In the 3d effective theory, the Lorentzian background geometry is 
determined by analytic continuation, for instance 
$\chi\to i\overline{\chi}+\pi/2$ (and $\xi\to\overline{\xi}$, 
$\rho\to\overline{\rho}$), which generates the line element
\beq
ds^2 = d\overline{\xi}^2+\overline{\rho}^2(\overline{\xi})
\left[-d\bc^2+\cosh^2(\bc)\,d\phi^2 \right] \,.
\label{Lmetric}
\eeq
This converts the point $\xi=0$ into a hypersurface:  the future 
lightcone of the nucleation event (at $\bc=0$).  The spacetime can 
be extended inside this lightcone by analytic continuation, taking 
$\overline{\chi}\to\chi-i\pi/2$, $\overline{\xi}\to i\xi$, and 
$\overline{\rho}\to i\rho$.  This creates an open FRW bubble interior.  
After returning the conformal factor $e^{-\Psi}=e^{-\alpha\psi}$ of 
the original metric ansatz (\ref{ansatz}), and revealing the compact 
dimension $z$, the internal bubble geometry has a line element of 
the form 
\beq
ds^2 = -e^{-\alpha\psi(\xi)}d\xi^2 + e^{-\alpha\psi(\xi)}
\rho^2(\xi)\left[d\chi^2+\sinh^2(\chi)\,d\phi^2\right] 
+ L^2e^{\alpha\psi(\xi)}\, dz^2 \,.
\label{bubblegeom0}
\eeq

\section{Background bubble evolution}
\label{sec:background}

In the previous section we found that bubble nucleation via 
modulus decay creates an internal bubble geometry corresponding 
to a (2+1)-dimensional, open FRW manifold crossed with an 
expanding circle, as given by (\ref{bubblegeom0}).  
Henceforth we use more traditional geometric variables, with 
respect to which the line element is written
\beq
ds^2 = -d\tau^2 + a^2(\tau)
\left[d\chi^2+\sinh^2(\chi)\,d\phi^2\right] 
+ b^2(\tau)\, dz^2 \,.
\label{bubblegeom}
\eeq 
We now see that the metric is of the Kantowski--Sachs form (case 
2 of \cite{KS}), a Bianchi type III homogeneous but anisotropic 
geometry.  

To understand the internal dynamics of bubbles like ours, we add 
to the Lagrangian some matter degrees of freedom.  These, we 
presume, have stress-energy tensor components that are all 
initially negligible next to those of the inflaton.  Furthermore, 
we presume the inflaton field (initially) reflects the symmetries 
of the tunneling instanton; in particular we take the inflaton to 
be static at the instanton boundary and independent of the 
initially-compact dimension $z$.  It then suffices to treat the 
matter as a perfect fluid, with stress-energy tensor 
$T^{\mu}_{\phantom{\mu}\nu}={\rm diag}\{-\rho,\,p,\,p,\,p\}$,
where the energy density $\rho$ and pressure density $p$ are 
(neglecting for the moment quantum fluctuations) functions of time 
alone.  The field equations are then
\bea
\frac{\dot{a}^2}{a^2} 
+ 2\frac{\dot{a}}{a}\frac{\dot{b}}{b} - \frac{1}{a^2}
&=& 8\pi G\Big[ V(b) + \rho\,\Big] \label{E2}\\
\frac{\ddot{a}}{a}-\frac{\ddot{b}}{b}
+\frac{\dot{a}}{a}\bigg(\frac{\dot{a}}{a}-\frac{\dot{b}}{b}\bigg)
-\frac{1}{a^2}
&=& 8\pi G\,b\, V'(b) \label{Eb}\\
\dot{\rho} + \bigg(2\frac{\dot{a}}{a} + 
\frac{\dot{b}}{b}\bigg)\left( \rho+p \right) 
&=& 0\,, \label{F2}
\eea  
where here and below the prime denotes differentiation with respect 
to $b$, the dot differentiation with respect to bubble proper time 
$\tau$, while the term $V(b)$ is described below.  In the second 
equation, (\ref{Eb}), we have combined two of the Einstein field 
equations in order to express a relation that will be useful later.

The anisotropy of the bubble is encoded in the initial conditions 
for bubble evolution, which are in turn established by the boundary 
conditions of the tunneling instanton.  These set the spatial and 
time derivatives of all non-metric fields to zero, so at $\tau=0$ 
we have  
\beq
\rho = \rho_{\rm inf}\,,\qquad \dot{\rho} = 0\,,\qquad 
p=-\rho_{\rm inf}\,, \qquad \dot{p}=0 \,,
\label{ic}
\eeq
where as noted above we assume the initial matter density is dominated 
by the inflaton.  Because the ``volume'' modulus of the $z$ dimension 
is the tunneling field, the instanton boundary conditions require that 
$b(\tau)$ approach a constant as $\tau\to0$, i.e.
\beq
 b \to L\,e^{\alpha\psi_d/2}\equiv b_d\,,\qquad \dot{b} \to 0\,.
\label{bsol}
\eeq
Meanwhile, referring to (\ref{boundary1}) and (\ref{boundary2}) 
and being careful to track the various variable redefinitions, we find 
that as $\tau\to 0$, 
\beq
a\to 0\,, \qquad \dot{a}\to 1\,.
\label{asol}
\eeq

In Section \ref{sec:bubble} it was convenient to absorb all of 
the matter degrees of freedom into the cosmological constant $\Lambda$.  
Moreover, in the 3d effective theory with the $z$ coordinate integrated 
out, $\Lambda$ coupled with the modulus $\psi$ in the 3d effective 
potential $\overline{V}(\psi)$.  Now we have returned to the 4d picture, 
in which $\Lambda$ is a bare cosmological constant, and have introduced 
$\rho$ and $p$ to generically characterize the matter content of the 
universe.  Therefore it is convenient to shift the cosmological constant 
$\Lambda$ into those quantities.  The other ingredients of the (toy) 
compactification model of Section \ref{ssec:compactification} generate 
a 4d ``potential'' for $b$,  
\beq
V(b) = \frac{1}{2}n^2\eta^2b^{-2} 
+ \frac{\kappa_2}{2}n^4\eta^4b^{-4}
+ \frac{\kappa_3}{2}n^6\eta^6b^{-6} \,,
\eeq 
where we note the change of variable $d\tau=e^{-\alpha\psi/2}\,dx^0$, 
which also affects the form of $V(b)$.  

For simplicity we treat the inflaton energy density as equivalent to 
vacuum energy density, $\rho_{\rm inf}=$ constant. (Thus, according to 
the above assumptions, $\rho_{\rm inf}$ is almost identical to 
$\Lambda$, the difference being the present dark energy.)  It is 
convenient to also assume the inflaton energy density dominates over 
the modulus potential, even right after bubble nucleation, i.e. 
$\rho_{\rm inf}\gg V(b_d)$.  This allows us to obtain the simple 
analytic solution,
\beq
a(\tau) = H_{\rm inf}^{-1}\sinh(H_{\rm inf}\,\tau) \,, \qquad
b(\tau) = b_d\cosh(H_{\rm inf}\,\tau) \,,
\label{sols}
\eeq
where $H_{\rm inf}^2\equiv 8\pi G\,\rho_{\rm inf}/3$.  (Note that the 
inequality $\rho_{\rm inf}\gg V(b_d)$ is satisfied by the choice of 
parameters used to generate Figure \ref{fig:V}.)  We have chosen the 
normalization of the scale factor $b(\tau)$ so as to conform to the 
initial condition given in (\ref{bsol}).    

One might notice that this solution resembles a slicing of 4d de 
Sitter space.  In fact, in the (4+1)-dimensional Minkowski 
($ds^2=-dt^2+du^2+dv^2+dx^2+dy^2$) embedding:      
\bea
t &=& H_{\rm inf}^{-1}\sinh(H_{\rm inf}\,\tau)\,
\cosh(\chi)\\
u &=& H_{\rm inf}^{-1}\cosh(H_{\rm inf}\,\tau)\,
\cos(b_dH_{\rm inf}\,z)\\
v &=& H_{\rm inf}^{-1}\cosh(H_{\rm inf}\,\tau)\,
\sin(b_dH_{\rm inf}\,z)\\
x &=& H_{\rm inf}^{-1}\sinh(H_{\rm inf}\,\tau)\,
\sinh(\chi)\,\cos(\phi) \\
y &=& H_{\rm inf}^{-1}\sinh(H_{\rm inf}\,\tau)\,
\sinh(\chi)\,\sin(\phi) \,,
\eea
the bubble coordinates $\{\tau,\,\chi,\,\phi,\,z\}$ sit on
the hyperboloid $-t^2+u^2+v^2+x^2+y^2=H_{\rm inf}^{-2}$, on 
which (\ref{sols}) gives the scale factors of the induced metric.  
However this is merely a curiosity; it arises only because we take 
the limit $\rho_{\rm inf}\gg V(b_d)$, with $\rho_{\rm inf}=$ const.  
Indeed, closer inspection reveals that that the Minkowski 
coordinates $u$ and $v$ are not periodic with the bubble 
coordinate $z$ (due to the factors of $b_dH_{\rm inf}$), 
indicating that the bubble geometry covers only a subset of the
hyperboloid, with periodic boundary conditions.\footnote{We can 
also define the time coordinate 
$\xi\equiv H_{\rm inf}^{-1}\sinh(H_{\rm inf}\,\tau)$, in which case 
the line element becomes
\beq
ds^2 = -\left(1+H_{\rm inf}^2\xi^2\right)^{-1}
d\xi^2 + \xi^2\left[ d\chi^2+\sinh^2(\chi)\,d\phi^2 \right] 
+ b_d^2\left(1+H_{\rm inf}^2\xi^2\right)dz^2 \,,
\eeq
which is also seen as 4d de Sitter space when $b_d^2=H_{\rm inf}^{-2}$.  
Notice that we can now read off the 3d dynamical fields of 
(\ref{bubblegeom0}):  
$\rho(\xi)=\xi\left(1+H_{\rm inf}^2\xi^2\right)^{1/2}$ and 
$\psi(\xi)=\alpha^{-1} \ln\!\left(1+H_{\rm inf}^2\xi^2\right)$, 
with $L=b_d$.}  

Note that the bubble geometry features two forms of anisotropy: 
expansion anisotropy---stemming from the two distinct scale 
factors $a$ and $b$---and spatial curvature anisotropy---stemming 
from the ($\chi$,~$\phi$) plane being open while the orthogonal 
$z$ direction is flat.  During inflation both forms of anisotropy
rapidly redshift away.  Although the presence of anisotropy makes 
choosing the definition of a scalar curvature parameter somewhat 
ambiguous, we find it convenient to write
\beq
\Omega_{\rm curv}\equiv \frac{1}{3\,a^2H^2} \,, 
\label{curvedef}
\eeq   
where $H\equiv\dot{a}/a$ and, following the assumptions above, here 
and below treat the effect of the modulus potential as negligible 
next to the above term.  Notice that the relationship between 
$\Omega_{\rm curv}$ and $1/a^2H^2$ involves a factor of 1/3 that is 
not present in the isotropic case.  Meanwhile, the expansion 
anisotropy can be parametrized by
\beq
h\equiv \frac{\dot{a}}{a} - \frac{\dot{b}}{b} \,.
\eeq  
In (\ref{Eb}) we have combined two of the Einstein field equations
so that we can now write
\beq
\dot{h} +\bigg(2\frac{\dot{a}}{a}+\frac{\dot{b}}{b}\bigg)h 
=\frac{1}{a^2} \,.
\eeq
Thus we see that anisotropic spatial curvature sources anisotropic 
expansion.  

While it would be interesting to understand any observable effects 
of late-time spatial curvature and expansion anisotropy (some of 
which have been explored in \cite{Thorne,CH,DLN,BJS,DD}), for the 
purpose of this paper we set such questions aside.  Indeed, to gain 
some understanding of the statistical anisotropy of inflationary 
perturbations in Section \ref{sec:perts}, we project the spectrum
onto a two-sphere ``of last-scattering,'' treating the background 
evolution as if the expansion anisotropy $h$ can be ignored.  As 
suggested above and shown in greater detail in \cite{DD}, this 
approach is not entirely self-consistent, which is to say the actual 
surface of last scattering is deformed by the presence of anisotropic 
expansion $h$.  However the analysis allows us to gain a qualitative 
understanding of some of the effects of anisotropic bubble 
nucleation; we leave the more complicated, complete treatment to 
future work.    

For the moment, we simply note two important relationships for 
later reference.  The first is the leading-order comoving distance 
to the surface of last scattering,  
\beq
\nr_\star = \int_{\tau_\star}^{\tau_0} \frac{d\tau}{a(\tau)}
\simeq \frac{3.5}{a_0H_0}
\simeq 6.1 \sqrt{\Omega_{\rm curv}^0}\leq 0.50 \,, 
\label{chitoday}
\eeq
where, to obtain the leading-order expression, we have ignored the 
effects of spatial curvature and background anisotropy on the 
evolution of $a$ (we have included the effect of dark energy, 
modeled as cosmological constant with $\Omega_\Lambda/\Omega_m=2.85$).  
Here a subscript ``0'' designates quantities evaluated at the point of
present detection while a subscript ``$\star$'' designates quantities 
evaluated at the surface of last scattering, and the last inequality 
follows from the observational limit 
$\Omega_{\rm curv}^0 \leq 6.6\times 10^{-3}$ 
(WMAP+BAO+SN with $w=-1$ prior, treating the 95\% confidence level as 
if it were a hard bound) \cite{WMAP}; however it should be emphasized 
that late-time anisotropic expansion induces a quadrupole in the CMB, 
itself appearing at order $\Omega^0_{\rm curv}$, which absent 
cancellations provides a stronger constraint on $\Omega^0_{\rm curv}$ 
\cite{DD}.    

The second important relationship comes from realizing that periodicity
in one of our large spatial dimensions has not been observed (see 
e.g.~\cite{O-CTZH}), indicating that the  physical circumference of the 
closed $z$ dimension must be greater than the physical diameter of the 
surface of last scattering, or  
\beq
2\pi\, b_\star \geq 2\,a_\star
\int_{\tau_\star}^{\tau_0} \frac{d\tau}{a(\tau)} \,.
\eeq
Ignoring the subleading effect of late-time anisotropic expansion,
we see from (\ref{sols}) that we can write 
$b_\star= b_dH_{\rm inf}\,a_\star$.  Inserting the result of the
last paragraph then gives the constraint
\beq
\Omega_{\rm curv}^0 \leq 0.27\,(b_dH_{\rm inf})^2 \,.
\label{curvbHconstraint}
\eeq
Thus, as the quantity $b_dH_{\rm inf}$ is decreased, the maximum allowed 
present-day curvature parameter is decreased, along with the size 
of all related effects.  However it should be noted that this 
result depends crucially on the ``redshift factor'' $b_dH_{\rm inf}$ 
between the two scale factors $a$ and $b$, which in turn relies on 
our limit of approximation to obtain the analytic solution 
(\ref{asol}).  It is not clear how this constraint is modified outside
of the limit $\rho_{\rm inf}\gg V(b_d)$.

\section{Inflationary perturbations}
\label{sec:perts}

\subsection{Power spectrum of a massless scalar}
\label{ssec:power}

We would like to avoid the formidable task of developing cosmological 
perturbation theory about the anisotropic background of 
(\ref{bubblegeom}).  Therefore, we assume that metric perturbations 
are negligible, until a period when the local background anisotropy 
is negligible, during which standard cosmological perturbation theory 
can be used.  This could very well be a generic feature of realistic 
models of inflation, since at least in isotropic models the metric 
perturbations are suppressed (by the first slow-roll parameter) 
relative to inflaton fluctuations, until their wavelengths have 
expanded far beyond the inflationary Hubble radius (see 
e.g.~\cite{LS}).  However to be concrete and self-consistent we focus 
on the fluctuations in a subdominant scalar field $\sigma$.  The 
fluctuations in $\sigma$ may be converted into adiabatic density 
perturbations by a variety of proposed mechanisms; for instance 
$\sigma$ could be a curvaton \cite{curvaton}.  Because we ignore 
metric perturbations, we must set aside the interesting question of 
what modifications to the spectrum of tensor perturbations one might 
observe in this scenario.
  
For simplicity we assume that during inflation within the bubble the 
mass of $\sigma$ is negligible ($m_\sigma\ll H_{\rm inf}$), as are the 
interactions between $\sigma$ and any other matter fields, so that the 
only terms involving $\sigma$ that appear in the Lagrangian are given 
by
\beq
S_{\sigma} = -\int\! \sqrt{-g}\,d^4x \,
\frac{1}{2}g^{\mu\nu}\partial_\mu\sigma\partial_\nu\sigma \,.
\label{L}
\eeq
The background metric $g_{\mu\nu}$ is given by (\ref{bubblegeom}); 
however we find it convenient here to work in terms of the 
``conformal'' time $\eta_c$, where
\beq
\eta_c\equiv \int \frac{d\tau}{a(\tau)} \,.
\label{ct}
\eeq
Note that we use the term ``conformal'' loosely, as this time 
parametrization does not make the metric entirely conformally flat.  
Instead, we have the line element
\beq
ds^2 = a^2(\eta_c)
\left[-d\eta_c^2 + d\chi^2+\sinh^2(\chi)\,d\phi^2\right] 
+b^2(\eta_c)\,dz^2\,.
\label{cmetric}
\eeq

We solve the equation of motion of $\sigma$ by separation of variables, 
writing
\beq
\sigma_{\upsilon qrs}(\eta_c,\,\chi,\,\phi,\,z) 
=\frac{1}{\sqrt{ab}}\,\Upsilon^\upsilon_{qs}(\eta_c)\,U_{qrs}(\bfx)
= \frac{1}{\sqrt{ab}}\,\Upsilon^\upsilon_{qs}(\eta_c)\,X_{qr}(\chi)\,
\Phi_r(\phi)\,Z_s(z)\,,
\label{bubmodedef}
\eeq
where the placement of indices anticipates results below.  The 
factor $(ab)^{-1/2}$ is introduced so that the temporal mode functions
$\Upsilon^\upsilon_{qs}$ are canonically normalized.  The resulting 
equations for $\Phi_r$ and $Z_s$ are very simple:
\beq
\Phi_r''(\phi) = -r^2\,\Phi_r(\phi)\,, \qquad 
Z_s''(z) = -s^2\,Z_s(z) \,,
\eeq
where here and below we use a prime to denote differentiation with 
respect to the lone argument of the function that it accents.  The 
factors $r^2$ and $s^2$ parametrize the separation of variables.  
A complete set of orthonormal solutions are 
\beq
\Phi_r(\phi) = \frac{1}{\sqrt{2\pi}}\, e^{ir\phi}\,, \qquad
Z_s(z) = \frac{1}{\sqrt{2\pi}}\, e^{is z} \,, 
\label{TZmodes}
\eeq
where the periodic boundary conditions on $\phi$ and $z$ dictate 
that $r$ and $s$ must be (positive or negative) integers (recall that 
we scale the $z$ direction so that it is periodic on $0\leq z\leq 2\pi$).  

Meanwhile, the differential equation for the $X_{qr}$ can be written
\beq
(1-c^2)\,X_{qr}''(c)-2c\,X_{qr}'(c) =
\left(\frac{1}{4}+q^2+\frac{r^2}{1-c^2}\right) X_{qr}(c) \,,
\label{Xeq}
\eeq
where the sum $1/4+q^2$ corresponds to the third separation constant 
in the above separation of variables.  We have defined the variable
$c=\cosh(\chi)$ so as to obtain the Legendre equation (as per our 
convention the primes now denote derivatives with respect to $c$), 
the solutions of which are the associated Legendre functions 
$P^r_{iq-1/2}$ and $Q^r_{iq-1/2}$ \cite{ASmath}.  We take interest in 
the solutions $P^r_{iq-1/2}$, with $q$ real and positive, which are 
finite and stationary in the limit $c\to 1$.  A convenient 
normalization is \cite{BI}
\beq
X_{qr}(c) = \frac{\Gamma(\frac{1}{2}+iq-r)}{\Gamma(iq)}\,
P^r_{iq-1/2}(c)\,,
\label{Xmodes}
\eeq
where $\Gamma$ denotes the gamma function.  The $X_{qr}$ then satisfy 
the orthonormality condition,
\beq
\int_1^\infty dc\, X_{qr}(c)\,X_{\q r}^*(c) = \delta(q-\q) \,,
\label{orthogonality}
\eeq
where $\delta$ denotes the Dirac delta function.  Functions of this sort
are studied in \cite{BI}, and it is straightforward to extend those
analyses to the present situation.  Using the shorthand $U_{qrs}(\bfx)$
to denote the combined spatial mode functions, one can show that
\beq
\sum_{r,\,s} U_{qrs}(\bfx_1)\,U_{qrs}^*(\bfx_2) = 
\frac{q\tanh(\pi q)}{4\pi^2}\sum_s 
P_{iq-1/2}(\Xi_{12})\,e^{is(z_1-z_2)} \,, 
\label{addition}
\eeq   
where $\Xi_{12}\equiv\cosh(\chi_1)\cosh(\chi_2)-\sinh(\chi_1)\sinh(\chi_2)
\cos(\phi_1-\phi_2)$.  By following the analogous calculation in 
\cite{BI}, this ``addition theorem'' can be used to prove the 
completeness relation
\beq
\int_0^\infty dq\,\sum_{r,\,s}\,
U_{qrs}(c_1,\,\phi_1,\,z_1)\,U_{qrs}^*(c_2,\,\phi_2,\,z_2) =
\delta(c_1-c_2)\,\delta(\phi_1-\phi_2)\,
\delta(z_1-z_2) \,.
\label{completeness}
\eeq

The time evolution of $\sigma$ is given by the mode functions 
$\Upsilon^\upsilon_{qs}$, which satisfy
\beq
\ddot{\Upsilon}^\upsilon_{qs}(\eta_c) 
= -\left(s^2\frac{a^2}{b^2} + \frac{1}{4} + q^2
-\frac{1}{2}\frac{\ddot{a}}{a}-\frac{1}{2}\frac{\dot{a}\dot{b}}{ab}
-\frac{1}{2}\frac{\ddot{b}}{b}+\frac{1}{4}\frac{\dot{a}^2}{a^2}
+\frac{1}{4}\frac{\dot{b}^2}{b^2}\right)
\Upsilon^\upsilon_{qs}(\eta_c) \,.
\label{Upsilon}
\eeq
To proceed we must specify the (conformal) time dependence of the 
scale factors $a$ and $b$.  For simplicity we adopt the analytic 
solution of Section \ref{sec:background}, given by 
(\ref{sols}), which treats the inflaton energy density as constant.  
In terms of this solution, the conformal time is 
$\eta_c=\ln\left[\tanh(H_{\rm inf}\,\tau/2)\right]$, which runs 
from minus infinity to zero as the bubble proper time $\tau$ runs 
from zero to infinity.  The two scale factors are then given by
\beq
a(\eta_c) = -H_{\rm inf}^{-1}\,{\rm csch}(\eta_c) \,,
\qquad
b(\eta_c) = -b_d\,{\rm coth}(\eta_c) \,.
\label{scalefactors}
\eeq 
Some simplification occurs when we plug this solution into 
(\ref{Upsilon}), and in the end we find
\beq
\ddot{\Upsilon}^\upsilon_{qs}(\eta_c) = -
\left[q^2 -2\,{\rm csch}^2(\eta_c)+\left(\mu^2-\frac{1}{4}\right)
{\rm sech}^2(\eta_c)\right]\Upsilon^\upsilon_{qs}(\eta_c) \,,
\label{upsiloneom}
\eeq
where we have defined $\mu\equiv s/b_dH_{\rm inf}$.  The solutions to 
this equation can be written 
\bea
\Upsilon^\upsilon_{qs}(\eta_c) &=&
-2^{-iq}C^\upsilon_1\coth(\eta_c) \cosh^{iq}(\eta_c)\, 
F\!\left[-\frac{1}{4}-\frac{iq}{2}-\frac{\mu}{2},\,
-\frac{1}{4}-\frac{iq}{2}+\frac{\mu}{2},\,
1-iq;\, {\rm sech}^2(\eta_c)\right] \nn\\
&& -2^{iq}C^\upsilon_2\coth(\eta_c)\,{\rm sech}^{iq}(\eta_c)\,
F\!\left[-\frac{1}{4}+\frac{iq}{2}-\frac{\mu}{2},\,
-\frac{1}{4}+\frac{iq}{2}+\frac{\mu}{2},\,
1+iq;\, {\rm sech}^2(\eta_c)\right] \qquad\,\,\,\label{ssol}\\ 
&\equiv& C^\upsilon_1 \, \cF_1(\eta_c) + C^\upsilon_2 \, \cF_2(\eta_c) 
\label{symsol}\,,
\eea
where $F\equiv\!\!\!\phantom{F}_2F_1$ denotes the hypergeometric 
function \cite{ASmath}.  The index $\upsilon$ is explained below.  
We have introduced the shorthand notations $\cF_1$ and $\cF_2$ for 
later convenience, and have chosen to represent the solutions so 
that, in the limit $\eta_c\to-\infty$, 
\beq
\cF_1(\eta_c) \,\to\, e^{-iq\eta_c} \,,\qquad
\cF_2(\eta_c) \,\to\, e^{iq\eta_c} \,. \label{F12lim1}
\eeq
For later reference we also note the asymptotic behavior as 
$\eta_c\to 0$, 
\bea
\frac{\cF_1(\eta_c)}{\sqrt{a(\eta_c)b(\eta_c)}} \,&\to&\, 
\sqrt{\frac{H_{\rm inf}}{b_d}}\,
\frac{2^{iq}\,\Gamma(\frac{3}{2})\,\Gamma(1-iq)}
{\Gamma(\frac{5}{4}-\frac{iq}{2}+\frac{\mu}{2})\,
\Gamma(\frac{5}{4}-\frac{iq}{2}-\frac{\mu}{2})} \label{F1lim2}\\
\frac{\cF_2(\eta_c)}{\sqrt{a(\eta_c)b(\eta_c)}} \,&\to&\, 
\sqrt{\frac{H_{\rm inf}}{b_d}}\,
\frac{2^{-iq}\,\Gamma(\frac{3}{2})\,\Gamma(1+iq)}
{\Gamma(\frac{5}{4}+\frac{iq}{2}+\frac{\mu}{2})\,
\Gamma(\frac{5}{4}+\frac{iq}{2}-\frac{\mu}{2})}\,.\label{F2lim2}
\eea

The initial conditions for the inflationary perturbations are set 
by the choice of integration constants $C^\upsilon_1$ and 
$C^\upsilon_2$.  Different choices of initial conditions correspond 
to different choices of the quantum vacuum state 
\cite{BirrellDavies}---a particularly attractive choice is that of 
the Bunch--Davies vacuum \cite{BunchDavies}, which maps de Sitter 
mode functions onto zero-occupation Minkowski wavefunctions in the 
small-scale limit.  Within the context of bubble nucleation in an 
inflating background, the Bunch--Davies state is determined by 
tracing the evolution of mode functions back into the parent vacuum 
and Klein--Gordon normalizing the positive-frequency modes on a 
Cauchy surface \cite{BGT,YST,GMST}.  The analysis is rather tedious,
and we relegate it to Appendix \ref{sec:BD}.  The results are inserted 
at an appropriate point below.  

Having solved for the mode functions of $\sigma$, to obtain the 
spectrum of inflationary perturbations we promote $\sigma$ to a 
quantum operator, $\hat{\sigma}$, and compute the two-point correlation 
function.  The analysis can proceed in exact analogy to the standard 
formalism (see e.g.~\cite{MFB}).  In particular, we can express 
$\hat{\sigma}$ as a mode expansion of creation and annihilation 
operators,
\beq
\hat{\sigma}(\eta_c,\,\bfx) = 
\int dq\,\sum_{\upsilon,\,r,\,s,}\, 
\frac{1}{\sqrt{a(\eta_c)\,b(\eta_c)}}\left[
\Upsilon^\upsilon_{qs}(\eta_c)\,U_{qrs}(\bfx)\,\hat{a}_{\upsilon qrs}
+\Upsilon^{\upsilon*}_{qs}(\eta_c)\,U_{qrs}^*(\bfx)\,
\hat{a}_{\upsilon qrs}^\dagger\right] \,,
\label{sigmaoperator}
\eeq
where $\hat{a}_{\upsilon qrs}$ and $\hat{a}^\dagger_{\upsilon qrs}$ 
satisfy the appropriate analogues of the standard commutation relations,
\beq
\big[\hat{a}_{\upsilon qrs},\,\hat{a}_{\upsilon'\q\r\s}^\dagger\big] 
= \delta(q-\q)\,\delta_{r\r}\,\delta_{s\s}\,\delta_{\upsilon\upsilon'}\,,
\quad \big[\hat{a}_{\upsilon qrs},\,\hat{a}_{\upsilon'\q\r\s}\big]
= \big[\hat{a}_{\upsilon qrs}^\dagger,\,
\hat{a}_{\upsilon'\q\r\s}^\dagger\big] = 0 \,,
\label{ladder}
\eeq
as do the canonical field operator $\sqrt{ab}\,\hat{\sigma}$ and its 
conjugate momentum field operator $\hat{\pi}$.  The index $\upsilon$
is explained in Appendix \ref{sec:BD}; it takes one of two values, 
which can be denoted $\pm$.  From (\ref{sigmaoperator}) we see that 
the ``Fourier'' transform of $\hat{\sigma}(\eta_c,\,\chi,\,\phi,\,z)$ 
can be written
\beq
\hat{\sigma}(\eta_c,\,q,\,r,\,s) = 
\frac{1}{\sqrt{a(\eta_c)b(\eta_c)}}\sum_{\upsilon}\left[
\Upsilon^\upsilon_{qs}(\eta_c)\,\hat{a}_{\upsilon qrs} - 
\Upsilon^{\upsilon*}_{(-q)(-s)}(\eta_c)\,
\hat{a}_{\upsilon (-q)(-r)(-s)}^\dagger \right]\,,
\label{conjugation}
\eeq
which gives the equal-time momentum-space two-point correlation 
function
\beq
\<\hat{\sigma}(\eta_c,\,q,\,r,\,s)\,
\hat{\sigma}^\dagger(\eta_c,\,\q,\,\r,\,\s)\>
= \frac{\sum_{\upsilon}|\Upsilon^\upsilon_{qs}(\eta_c)|^2}
{a(\eta_c)b(\eta_c)}\delta(q-\q)\,\delta_{r\r}\,\delta_{s\s} \,.
\label{power}
\eeq

Each ``Fourier'' mode $U_{qrs}$ associates the set of separation 
constants $\{q,\,r,\,s\}$ with a set of comoving distance scales 
in the $\chi$, $\phi$, and $z$ directions, according to the 
characteristic scales of variation of $U_{qrs}(\chi,\,\phi,\,z)$.  
After these comoving scales grow larger than the Hubble radius, the 
mode ``amplitude'' $\sum_\upsilon|\Upsilon^\upsilon_{qs}|^2/ab$ 
rapidly asymptotes to an $\eta_c$-independent constant.  Because of 
this, and because all observable scales in our universe first 
expanded beyond the Hubble radius deep in the inflationary epoch, 
for practical purposes we can safely evaluate the two-point 
correlator (\ref{power}) in the limit $\eta_c\to 0$.  This gives 
the power spectrum  
\bea
P_{qs} &\equiv& \lim_{\eta_c\to 0}\, 
\frac{1}{a(\eta_c)b(\eta_c)}
\sum_{\upsilon}|\Upsilon_{qs}^\upsilon(\eta_c)|^2 \\
&=& \frac{\pi H_{\rm inf}}{8\,b_d\sinh^2(\pi q)}
\left\{\pi\cosh(\pi q)
\left|\Gamma\!\left(\frac{5}{4}+\frac{iq}{2}+\frac{\mu}{2}\right)\right|^{-2}\,
\left|\Gamma\!\left(\frac{5}{4}+\frac{iq}{2}-\frac{\mu}{2}\right)\right|^{-2}
\right.\nn\\
&& -\left. \cos(\pi\sqrt{1-\mu^2})
{\rm Re}\!\left[\frac{2^{2iq}\,\Gamma(\frac{1}{2}-iq-\sqrt{1-\mu^2})\,
\Gamma(\frac{1}{2}-iq+\sqrt{1-\mu^2})}
{\Gamma^2(\frac{5}{4}-\frac{iq}{2}-\frac{\mu}{2})\,
\Gamma^2(\frac{5}{4}-\frac{iq}{2}+\frac{\mu}{2})}\right]\right\} 
\label{mainpower}\qquad\\
&\approx& \frac{H_{\rm inf}^2}{2\,b_dH_{\rm inf}}
\left(q^2+\mu^2\right)^{-3/2} \,, \label{estimate}
\eea
where we have inserted the results from Appendix \ref{sec:BD}.  The last 
relation corresponds to taking the asymptotic limit of large $q$ 
and $\mu$.\footnote{Instead of performing the involved
analysis of Appendix \ref{sec:BD}, one might determine the integration 
constants $C^\upsilon_1$ and $C^\upsilon_2$ by studying the behavior of 
$\Upsilon^\upsilon_{qs}$ at very early bubble times, $\eta_c\to-\infty$, 
and on very small bubble scales, $q\to\infty$, and equating it with 
that of a free field in Minkowski space, 
$\Upsilon^\upsilon_{qs}\to (2q)^{-1/2}e^{-iq\eta_c}$.
(Note that the equation specifying $\Upsilon^\upsilon_{qs}$ takes the 
form $\ddot{\Upsilon}^\upsilon_{qs} = -q^2\,\Upsilon^\upsilon_{qs}$ at 
early conformal times---hence the ``Minkowksi'' wavefunction is 
independent of $s$.)  This corresponds to choosing the analogue of 
the so-called ``conformal'' vacuum, see e.g.~\cite{BirrellDavies}.  
It has the significant drawback of predicting an energy density of 
fluctuations that diverges as $\eta_c\to -\infty$, thus converting 
the mere coordinate singularity at the instanton boundary into a 
physical singularity.  Proceeding nevertheless, in this situation one 
finds no cause for the index $\upsilon$, and inspecting the asymptotic 
behavior of $\cF_1$ and $\cF_2$ in (\ref{F12lim1}) it is apparent that 
one would choose $C_1=(2q)^{-1/2}$ and $C_2=0$.  The resulting power 
spectrum is given by the first term of (\ref{mainpower}), but divided 
by $\coth(\pi q)$.  It approaches the same limiting behavior as the 
Bunch--Davies vacuum, (\ref{estimate}).}

To begin to understand this result, first note that the separation 
constant $q$ characterizes comoving radial distance scales in the 
($\chi$,~$\phi$) plane, and thus relates to the standard Cartesian 
wavenumbers $k_x$ and $k_y$ via $q^2 \sim k_x^2+k_y^2$ in the flat 
space (large $q$) limit.  Meanwhile, physical scales in the $z$ 
direction are redshifted relative to those in the ($\chi$,~$\phi$) 
plane due to the dissimilar evolution of the scale factors $a$ and 
$b$ at early times.  The late-time effect is an additional factor of 
$b(\eta_c)/a(\eta_c)|_{\eta_c\to0}=b_dH_{\rm inf}$ relating physical 
and comoving distances; therefore we relate the separation constant 
$s$ to the late-time Cartesian comoving wavenumber $k_z$ via 
$s\sim b_dH_{\rm inf}\,k_z$.  Thus the term in parentheses in 
(\ref{estimate}) approaches $(k_x^2+k_y^2+k_z^2)^{-3/2}$, which gives
the standard scale-invariant power spectrum (the factor of 
$1/b_dH_{\rm inf}$ will cancel when integrating over wavenumbers to
compute observables in the isotropic limit).  

Although we have just remarked on the congruence between the 
asymptotic limit of our result, (\ref{estimate}), and the result from 
standard, flat, isotropic inflation, the two inflationary spectra are 
not the same.  The power spectrum (\ref{estimate}) is expressed in 
terms of the anisotropic ``Fourier'' modes $U_{qrs}(\bfx)$, while for 
instance the isotropic Cartesian Fourier modes are $e^{i\bfk\cdot\bfy}$, 
where $\bfy=\{y_1,\,y_2,\,y_3\}$ are Cartesian coordinates.  This 
difference affects the observed spectrum because the (lack of) 
correlations implied by the delta functions 
$\delta(q-q')\,\delta_{rr'}\,\delta_{ss'}$ in the two-point correlator
(\ref{power}) are different than the (lack of) correlations implied by 
$\delta^{(3)}(\bfk-\bfk')$ in the standard isotropic case.

\subsection{Projection onto spherical harmonics}
\label{ssec:CMB}

At present the spectrum of primordial density perturbations is most
tightly constrained by measurements of fluctuations in the temperature
of photons streaming from the surface of last scattering, i.e.~CMB
fluctuations.  It is conventional to use spherical harmonics to cover 
a two-sphere representing our field of vision (here parametrized by 
angular coordinates $0\leq\theta\leq\pi$ and $0\leq\phi\leq 2\pi$).  
The observables are therefore the multipole moments
\beq
a_{\ell m} = \int d\zeta\,d\phi\,\, 
Y_{\ell m}^*(\zeta,\,\phi)\,
\delta_T(\zeta,\,\phi) \,,
\eeq
where $\delta_T$ represents the temperature fluctuation and for 
convenience we have defined $\zeta\equiv\cos(\theta)$.  The
(orthonormal) spherical harmonics are given by
\beq
Y_{\ell m}(\zeta,\,\phi) = 
\sqrt{\frac{(2\ell+1)(\ell-m)!}{4\pi\,(\ell+m)!}}\,
P^m_\ell(\zeta)\,e^{im\phi} \,.
\label{sphericalharmonics}
\eeq

The temperature fluctuation $\delta_T$ is sourced by density 
perturbations on the surface of last scattering, but the observed
CMB perturbations include secondary effects, which are incurred as 
photons stream from last scattering to the point of detection.  For 
simplicity we here ignore these effects.  In fact, we completely ignore 
the presence of late-time anisotropic expansion 
$h\equiv \dot{a}/a-\dot{b}/b$.  In this case, null geodesics radiating 
from the origin of coordinates see a flat metric, equivalent to that of 
a cylindrical coordinate system.  Meanwhile the corresponding flat 
cylindrical coordinates can be related to flat spherical coordinates, 
in terms of which the surface of last scattering has fixed radius 
$\nr_\star$ (again, neglecting $h$).  Thus we obtain     
\beq
\chi_\star(\zeta,\,\phi) = \nr_\star\sqrt{1-\zeta^2}\,,\qquad
\phi_\star(\zeta,\,\phi)= \phi\,,\qquad
z_\star(\zeta,\,\phi) = \frac{\nr_\star\,\zeta}{b_dH_{\rm inf}}\,,
\label{coordmatching}
\eeq   
which reflects a particular choice of matching between the  
cylindrical and the spherical coordinates (recall that 
$\zeta=\cos(\theta)$).  The factor $1/b_dH_{\rm inf}$ in $z$ 
comes from the ratio of scale factors $a$ and $b$ when matching 
onto the isotropic cylindrical coordinates.

We emphasize that while this procedure allows to create a picture of 
the inflationary spectrum in terms of spherical harmonics, the 
actual observed spectrum will contain corrections, coming from the 
anisotropic expansion between the surface of last scattering and the 
point of present detection.  In particular, the presence of 
anisotropic expansion deforms the surface of last scattering away 
from the surface defined by (\ref{coordmatching}), and perturbs the
trajectories of geodesics as they radiate away from the point of 
observation.  Our analysis can be viewed as a first step toward 
understanding the inflationary spectrum; a more complete analysis 
of the observational signatures is left to future work.

To be precise, we simply take the temperature fluctuations to be 
given by
\beq
\delta_T(\zeta,\,\phi)\propto\sigma(\eta_c=0,\,
\bfx_\star(\zeta,\,\phi))\,,
\label{isotoadiabatic}
\eeq
where $\bfx_\star=\{\nr_\star,\,\zeta,\,\phi\}$ designates the 
coordinates of the surface defined by (\ref{coordmatching}).  Here we 
have assumed that the isocurvature fluctuations in the light scalar field 
$\sigma$ ultimately directly source the adiabatic CMB perturbations, 
and have ignored a model-dependent proportionality constant.  Also, 
in addition to ignoring the expansion anisotropy $h$, we have ignored 
the evolution of primordial perturbations after they enter the Hubble 
radius.

We are actually interested in the multipole correlation function 
$C_{\ell\ell'mm'}=\<\hat{a}^{\phantom{\dagger}}_{\ell m}
\hat{a}_{\ell' m'}^\dagger\>$.  Expanding 
$\sigma(\eta_c=0,\,\bfx_\star(\zeta,\,\phi))$ in terms of mode 
functions $U_{qrs}$ as in (\ref{sigmaoperator}), we obtain
\beq
C_{\ell\ell'mm'} = 
\int d\zeta_1\,d\zeta_2\,d\phi_1\,d\phi_2\,dq\sum_{r,\,s} P_{qs}\,
Y_{\ell m}^*(\zeta_1,\,\phi_1)\,Y_{\ell' m'}(\zeta_2,\,\phi_2)\,
U_{qrs}(\zeta_1,\,\phi_1)\,U_{qrs}^*(\zeta_2,\,\phi_2) \,, 
\label{C1}
\eeq  
where $P_{qs}$ is the power spectrum, given by (\ref{mainpower}), 
and  
\beq
U_{qrs}(\zeta,\,\phi) = 
\frac{\Gamma(\frac{1}{2}+iq-r)}{2\pi\, \Gamma(iq)}\,
P^r_{iq-1/2}\left[\cosh\!\sqrt{\nr_\star^2\,(1-\zeta^2)}\right]
e^{ir\phi+i\mu\,\nr_\star\zeta} \,,
\label{Uspherical}
\eeq
where $\mu=s/b_dH_{\rm inf}$.  Plugging the $Y_{\ell m}$ 
and $U_{qrs}$ into (\ref{C1}), we can immediately perform the 
integrations over $\phi_1$ and $\phi_2$.  This gives
\bea
C_{\ell\ell'mm'} &=& \delta_{mm'}
\sqrt{\frac{(2\ell+1)(2\ell'+1)(\ell-m)!(\ell'-m)!}
{16\pi^2\,(\ell+m)!(\ell'+m)!}} \int d\zeta_1\,d\zeta_2\,dq\,\sum_{s}\,
P_{qs} \nn\\
&& \!\times\, \bigg|\frac{\Gamma(\frac{1}{2}+iq-m)}{\Gamma(iq)}\bigg|^2
P^m_{iq-1/2}\!\left[\cosh\!\sqrt{\nr_\star^2(1-\zeta_1^2)}\right]
P^{m*}_{iq-1/2}\!\left[\cosh\!\sqrt{\nr_\star^2(1-\zeta_2^2)}\right]
\nn\\
&& \!\times\, {P^m_\ell}^*(\zeta_1)\, P^m_{\ell'}(\zeta_2)\,
e^{i\mu\nr_\star(\zeta_1-\zeta_2)} \,. \phantom{\bigg|\bigg|^2}
\label{C2}
\eea 
Because of the complexity of this result, we here limit our attention
to two basic goals:  demonstrating that, as expected, the 
$C_{\ell\ell'mm'}$ approach the standard, isotropic results in the 
limit of many $e$-folds of inflation, and understanding the qualitative 
features of the low-multipole $C_{\ell\ell'mm'}$ via approximate 
numerical integrations.  

The physical distance to the surface of last scattering is fixed by the 
late-time big bang expansion history.  To model this, we treat the 
physical distance from the origin to the surface defined by 
(\ref{coordmatching}) as fixed, $a_\star\nr_\star=$ constant, so that 
$\nr_\star$ decreases as the duration of slow-roll inflation increases.  
Thus, the limit of long-duration inflation corresponds to the limit of 
very small $\nr_\star$.  In this limit only modes with large wavenumbers 
$q$ are relevant to observation, and the normalized Legendre functions 
approach normalized Bessel functions,
\beq
\frac{\Gamma(\frac{1}{2}+iq-r)}{\Gamma(iq)}\,
P^r_{iq-1/2}\left[\cosh\!\sqrt{\nr_\star^2\,(1-\zeta^2)}\right]
\to \sqrt{q}\,J_r\!\left(\nr_\star q\sqrt{1-\zeta^2}\right) \,,
\label{PJlimit}
\eeq  
where to be clear $J_r$ is the $r$th-order Bessel function of the first 
kind \cite{ASmath}, and we ignore the unimportant overall phase.  (This 
result is most easily obtained by taking the appropriate limit of the 
underlying differential equation.)  With this substitution the 
angular integrations over $\zeta_1$ and $\zeta_2$ can be performed 
analytically, using a convenient mathematical equality \cite{convenient}:
\beq
\int d\zeta\, e^{i\gamma\cos(\alpha)\,\zeta}\,P^m_\ell(\zeta)\,
J_m\!\left[\gamma\sin(\alpha)\sqrt{1-\zeta^2}\right] = 
2i^{\ell-m}\,P^m_\ell[\cos(\alpha)]\,\, j_\ell(\gamma) \,,
\label{conv}
\eeq
where $j_\ell$ denotes the spherical Bessel function of order 
$\ell$ \cite{ASmath}.  This result holds for positive and 
negative $m$, and any $0\leq\alpha\leq\pi$.  Since 
$P^{m*}_\ell=(-1)^{-m}P^{m}_\ell$, this equality also holds if we 
replace the Legendre polynomial with its complex conjugate on both 
sides.  Note that it is always possible to choose the $\alpha$ and 
$\gamma$ of (\ref{conv}) so that 
$\gamma\cos(\alpha)=\pm \mu\,\nr_\star$ and 
$\gamma\sin(\alpha)=q\nr_\star$, in particular one sets  
\beq
\gamma=\nr_\star\sqrt{q^2+\mu^2} \,, \qquad 
\cos(\alpha) = \frac{\pm \mu}{\sqrt{q^2+\mu^2}} \,.
\eeq 
Putting all of this together, we obtain 
\bea
C_{\ell\ell'mm'} &=& \delta_{mm'}
\sqrt{\frac{(2\ell+1)(2\ell'+1)(\ell-m)!(\ell'-m)!}
{\pi^2\,(\ell+m)!(\ell'+m)!}}\,i^{\ell-\ell'}\!\! \int dq\,\sum_s\,
q\, P_{qs}\,\nn\\
& & \!\times\, j_\ell\!\left(\nr_\star\sqrt{q^2+\mu^2}\right) 
j_{\ell'}\!\left(\nr_\star\sqrt{q^2+\mu^2}\right)
P^{m*}_\ell\bigg(\frac{\mu}{\sqrt{q^2+\mu^2}}\bigg)
P^m_{\ell'}\bigg(\frac{\mu}{\sqrt{q^2+\mu^2}}\bigg) \,. \quad\,\,
\label{C3}
\eea

Our next approximation is to replace the sum over integers $s$ with an 
integral over real $\mu=s/b_dH_{\rm inf}$.  Demonstrating the strict 
validity of this approximation is tedious; however intuitively we 
expect it to be accurate at least insofar as 
$\nr_\star\ll b_dH_{\rm inf}$.  This is because modes with wavelength 
$\lambda\sim 1/s$ should not contribute significantly to observables 
on scales $\nr\ll\lambda$, and at the same time the discrete spectrum 
should be well-approximated by a continuum when $s\gg 1$, or 
$\mu\gg 1/b_dH_{\rm inf}$.  Converting the sum over $s$ into an 
integral allows us to make use of a convenient variable transformation,  
defining $k$ and $\Theta$ according to 
\beq
q=(k/\nr_\star)\,\sin(\Theta)\,, \qquad \mu=(k/\nr_\star)\,\cos(\Theta) \,,
\eeq
where $k\geq 0$ and $0\leq\Theta\leq \pi$.  
The Jacobian of the transformation gives 
$dq\,d\mu=\nr_\star^{-2}\,k\,dk\,d\Theta
=\nr_\star^{-2}\,[k/\sin(\Theta)]\,dk\,d\cos(\Theta)$, so that  
$C_{\ell\ell'mm'}$ can be written
\bea
C_{\ell\ell'mm'} &=& \delta_{mm'} 
\sqrt{\frac{(2\ell+1)(2\ell'+1)(\ell-m)!(\ell'-m)!}
{\pi^2\,(\ell+m)!(\ell'+m)!}}\, i^{\ell-\ell'}\,b_dH_{\rm inf}
\int dk\,d\cos(\Theta)\,\,
\frac{k^2}{\nr_\star^3}\,P_{qs}(k,\,\Theta) \nn\\
& & \!\times\, 
j_\ell(k)\,j_{\ell'}(k)\, P^{m*}_\ell\big[\cos(\Theta)\big]\, 
P^m_{\ell'}\big[\cos(\Theta)\big]\,.
\phantom{\bigg|\bigg|}
\label{C4}
\eea
After this variable redefinition one can also show that in the limit of 
small $\nr_\star$, the power spectrum approaches 
$P_{qs}(k,\,\Theta)\to (H_{\rm inf}/2b_d)\,\nr_\star^3/k^3$. The 
integrals over $\cos(\Theta)$ and $k$ can then be performed, giving the 
standard flat, isotropic result:
\bea
C^{(0)}_{\ell\ell'mm'} \equiv \lim_{\nr_\star\to 0} C_{\ell\ell'mm'}
= \frac{H_{\rm inf}^2}{2\pi\,\ell(\ell+1)}\,
\delta_{\ell\ell'}\,\delta_{mm'} \,.
\label{isotropicpart}
\eea 

Now let us turn to computing some of the low-multipole components
of $C_{\ell\ell'mm'}$.  The direct numerical estimation of (\ref{C2})
converges very slowly, given our limited computational resources.  
Nevertheless, we can proceed as above, replacing the sum over $s$ 
with an integral over $\mu$.  This by itself does not improve the
convergence of the numerical evaluation, however if we now replace 
the power spectrum $P_{qs}$ of (\ref{mainpower}) with its asymptotic 
limit (\ref{estimate}), the integral over $\mu$ can be performed 
analytically,
\beq
\int^{\infty}_{-\infty} d\mu 
\left(q^2+\mu^2\right)^{\!-3/2}
e^{i\mu\nr_\star(\zeta_1-\zeta_2)} =  
\frac{2\nr_\star|\zeta_1-\zeta_2|}{q}\,
K_1\left(\nr_\star q|\zeta_1-\zeta_2|\right) \,,
\eeq    
where $K_1$ is the (first order) modified Bessel function of the 
second kind.  

As explained above, we expect replacing the sum over $s$ with an 
integral over $\mu$ to be accurate when $\nr_\star\ll b_dH_{\rm inf}$.
Meanwhile these quantities are observationally constrained by 
(\ref{chitoday}) and (\ref{curvbHconstraint}), which combine to give 
$\nr_\star\leq 3.2\,b_dH_{\rm inf}$.  Thus there is an interesting
region in the parameter space that cannot be probed by this 
approximation.  On the other hand, the asymptotic limit of the power 
spectrum $P_{qs}$, (\ref{estimate}), is approached rapidly as $q$ 
increases; see Figure \ref{fig:spectrum}.  We have computed a few 
elements of $C_{\ell\ell'mm'}$ without this approximation, using 
$b_dH_{\rm inf}=1$ and $\nr_\star=0.5$, and the two agree at about 
the 10\% level.  Note that, for fixed $b_dH_{\rm inf}$, the accuracy 
of these approximations improves as $\nr_\star$ is decreased.
 
\begin{table}
\begin{center}
{\small
\begin{tabular}{|c||c|c|c|c|c|c|c|c|}
\hline
& \phantom{1}$\ell=2$\phantom{1} & \phantom{1}$\ell=3$\phantom{1} & 
\phantom{1}$\ell=4$\phantom{1} & \phantom{1}$\ell=5$\phantom{1} 
& \phantom{1}$\ell=6$\phantom{1} & \phantom{1}$\ell=7\phantom{1}$ & 
\phantom{1}$\ell=8$\phantom{1} & \phantom{1}$\ell=9$\phantom{1} \\
\hline\hline
$\ell'=\ell \phantom{+21}$ & $-1.2$ & $-1.1$ & $-1.0$ & $-1.0$ & $-1.0$ & 
$-1.0$ & $-1.0$ & -1.0 \\
\hline
$\ell'=\ell\pm 1$ & $0$ & $0$ & $0$ & $0$ & $0$ & $0$ & $0$ & $0$ \\
\hline
$\ell'=\ell\pm 2$ & $0.33$ & $0.35$ & $0.36$ & $0.37$ & $0.38$ & 0.40 & 
0.40 & 0.40 \\
\hline
\end{tabular}}
\caption{\label{table:1}The multipole correlator contrast 
$\delta C_{\ell\ell'mm'}$ for several values of $\ell$ and $\ell'$, 
$m=m'=0$, in units of $\Omega^{(0)}_{\rm curv}$, using 
$\nr_\star=0.01$ (see the main text for details).}
\end{center}
\end{table}

Putting everything together, our final expression for the 
$C_{\ell\ell'mm'}$ is
\bea
C_{\ell\ell'mm'} &=& \delta_{mm'}
\sqrt{\frac{(2\ell+1)(2\ell'+1)(\ell-m)!(\ell'-m)!}
{16\pi^2\,(\ell+m)!(\ell'+m)!}}\, \nr_\star H_{\rm inf}^2\! 
\int d\zeta_1\,d\zeta_2\,dq\,\,\frac{|\zeta_1-\zeta_2|}{q} \nn\\
&& \!\times\, \bigg|\frac{\Gamma(\frac{1}{2}+iq-m)}{\Gamma(iq)}\bigg|^2
P^m_{iq-1/2}\!\left[\cosh\!\sqrt{\nr_\star^2(1-\zeta_1^2)}\right]
P^{m*}_{iq-1/2}\!\left[\cosh\!\sqrt{\nr_\star^2(1-\zeta_2^2)}\right]
\nn\\
&& \!\times\, {P^m_\ell}^*(\zeta_1)\, P^m_{\ell'}(\zeta_2)\,
K_1\left(\nr_\star q|\zeta_1-\zeta_2|\right) \,. \phantom{\bigg|\bigg|^2}
\label{m&m,2}
\eea 
To understand the deviations from isotropy, it is convenient to 
define the quantity
\beq
\delta C_{\ell\ell'mm'}\equiv 
\frac{C_{\ell\ell'mm'}-C^{(0)}_{\ell\ell'mm'}}
{\max\{C^{(0)}_{\ell\ell mm},\,C^{(0)}_{\ell'\ell' m'm'}\}} \,,
\label{anisotropicpart}
\eeq
which gives the correction to the ``isotropic background'' 
$C^{(0)}_{\ell\ell mm}$ of (\ref{isotropicpart}), in units of a relevant 
component of $C^{(0)}_{\ell\ell mm}$.  Note that within the context of 
the above approximations, the only observable that enters 
$\delta C_{\ell\ell'mm'}$ is the comoving distance $\nr_\star$.  
Comparing numerical results for several different values of $\nr_\star$, 
we find (roughly) $\delta C_{\ell\ell'mm'}\propto\nr_\star^2$ when 
$\nr_\star\lesssim 1$.  Meanwhile (\ref{chitoday}) gives 
$\nr_\star^2\simeq 37\Omega_{\rm curv}^0$; thus we conclude the non-zero
components of $\delta C_{\ell\ell'mm'}$ scale with the present day 
curvature parameter $\Omega_{\rm curv}^0$.  In Table \ref{table:1} we 
display $\delta C_{\ell\ell'mm'}$, in units of $\Omega^{(0)}_{\rm curv}$, 
for several values of $\ell$ and $\ell'$, choosing for simplicity 
$m'=m=0$, and setting $\nr_\star=0.01$.  

The most salient feature of Table \ref{table:1} is the existence of 
off-diagonal terms with respect to $\ell$ and $\ell'$.  This is 
qualitatively not unlike the results of \cite{ACW}, in which the 
phenomenology of a Bianchi type I anisotropic universe was studied.  As 
was the case in that scenario, we find that the off-diagonal terms of 
$C_{\ell\ell'mm'}$ do not fall off very rapidly relative to relevant 
components of the background $C^{(0)}_{\ell\ell'mm'}$.  Indeed, the 
sequence of $\delta C_{\ell\ell'mm'}$ with $\ell'=\ell\pm 2$ appears to 
increase with increasing $\ell$.  Although computing 
$\delta C_{\ell\ell'mm'}$ at large $\ell'=\ell+2$ is extremely time 
consuming, we have explored several values of $\ell'$ up to $\ell'=20$ 
and observed that the sequence asymptotes toward a small constant 
value, about $0.4$ (in units of $\Omega^{(0)}_{\rm curv}$).  We 
have not displayed components of $\delta C_{\ell\ell'mm'}$ with 
$\ell'-\ell>1$ and odd, which we expect, like $\ell'-\ell=1$, to give 
precisely zero.  The components with $\ell'-\ell>2$ and even feature 
very slow numerical convergence, but also appear consistent with zero.

\begin{figure}[t!]
\begin{center}
\begin{tabular}{cc}
\includegraphics[width=0.475\textwidth]{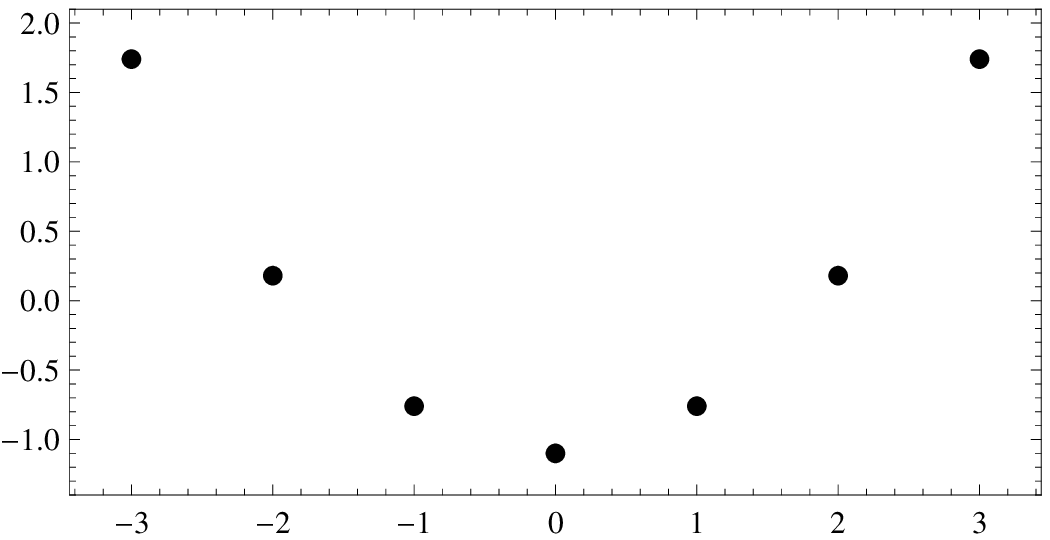} &
\includegraphics[width=0.475\textwidth]{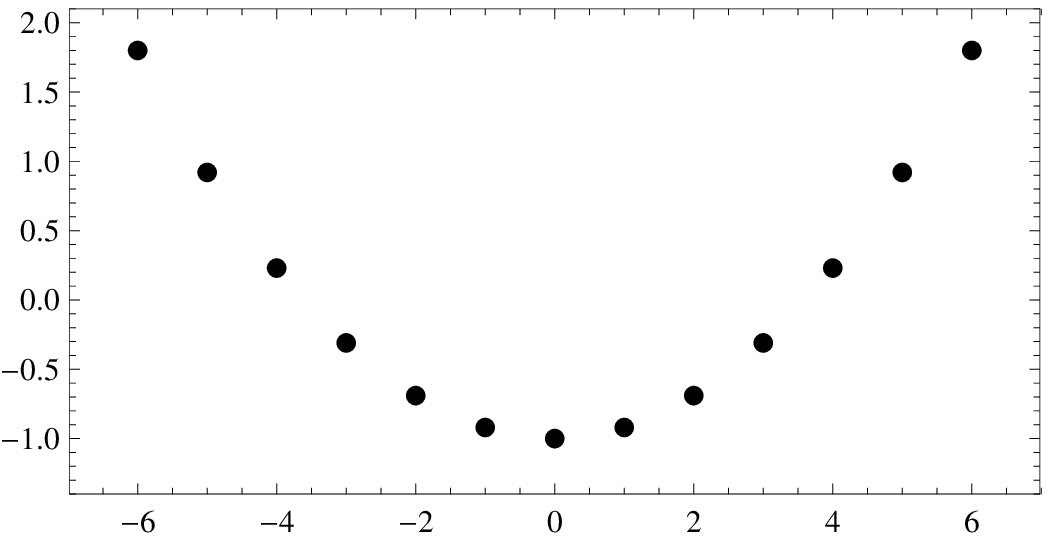} \\
\end{tabular}
\caption{\label{fig:m} Non-zero values of the the multipole correlator 
contrast $\delta C_{\ell\ell'mm'}$ for $\ell'=\ell=3,\,6$, in units 
of $\Omega^{(0)}_{\rm curv}$, using $\nr_\star=0.01$ (see the main 
text for details).}
\end{center}
\end{figure}

The failure of off-diagonal terms of $\delta C_{\ell\ell'mm'}$ to 
approach zero in the limit of large $\ell$ may at first seem to 
conflict with the usual interpretation that increasing multipole
moments $\ell$ correspond to probing smaller physical scales.  To 
understand this effect, we must recall that we are evaluating the
anisotropic modes on a surface that sits at fixed, non-zero 
$\nr_\star$.  The asymptotic behavior of 
(\ref{PJlimit}), in which the anisotropic modes converge to isotropic 
ones, corresponds to the limit $q\to\infty$, keeping $\nr_\star q$ 
constant, which must not be confused with the limit $q\to\infty$,
keeping $\nr_\star$ constant.  In the latter case, there is always 
a region on the two-sphere (corresponding to large intersection
with the open plane) over which there is a discrepancy between the 
anisotropic and isotropic modes, even in the limit $q\to\infty$.  
Since the spherical harmonics receive support over the entire 
two-sphere (regardless of $\ell$), this discrepancy corresponds to
statistical anisotropy in the projection onto spherical harmonics.  
We emphasize that the above comments concern the ratio that appears
in $\delta C_{\ell\ell'mm'}$; the observed multipole moments, 
$C_{\ell\ell'mm'}$, indeed decrease at increasing $\ell$ and $\ell'$.

Table \ref{table:1} displays results only for when $m'=m=0$.  To 
illustrate some of the dependence of $\delta C_{\ell\ell'mm'}$ on
the indices $m$ and $m'$, in Figure \ref{fig:m} we plot all of the 
non-zero entries of $\delta C_{\ell\ell'mm'}$ for $\ell'=\ell=3$ and 
$\ell'=\ell=6$.  The qualitative manifestation of statistical anisotropy
is evident:  at fixed $\ell'=\ell$, the correction 
$\delta C_{\ell\ell'mm'}$ increases with increasing magnitude of $m'=m$.  
The effect is on the same order as the $m'=m=0$ correction to 
$\delta C^{(0)}_{\ell\ell'mm'}$.   Recall that in standard, isotropic 
inflation the multipole correlator is independent of $m$ and $m'$.

Note that because the elements of $\delta C_{\ell\ell'mm'}$ do not fall 
off very rapidly with increasing $\ell$, there is hope to gather 
sufficient statistics to detect them, despite their being suppressed 
by roughly the size of the present-day curvature parameter; see 
e.g.~\cite{PK}.  While we consider it very interesting to explore what 
are the precise limits placed by cosmic variance on detecting these 
effects, such an investigation is beyond the scope of the present work.

\section{Plausibility}
\label{sec:likelihood}

The scenario described in this paper involves the convergence of a 
number of hypothetical circumstances.  Yet many of these ingredients 
are not crucial to the general idea, which we expect to cover a broad
set of models.  Furthermore, while the basic framework---bubble 
nucleation via modulus destabilization followed by not much more 
inflation than is necessary to flatten our Hubble volume---is 
certainly speculative, we do not consider it extremely implausible.  
We here take a moment to justify these attitudes.

An essential ingredient of our model is that the vacuum in which our 
bubble nucleates has one fewer large spatial dimensions than that of
our universe.  A toy model to implement this is described in 
Section~\ref{ssec:compactification}; however we consider this model 
only as a simple illustration of the concept.  If we accept ten- or 
eleven-dimensional string theory as a fundamental description of 
nature, then the existence of our (3+1)-dimensional pocket is proof 
of the compactification principle, and it seems reasonable to 
presume that other compactifications are possible and that quantum 
transitions among them occur during eternal inflation.

The mere occurrence of such transitions does not imply that one is 
likely to be the progenitor of our universe.  One might argue that, 
all else being equal, the nucleation of our bubble in a vacuum with 
a greater number of large (expanding) spatial dimensions is more 
likely, because of the greater physical volume available to nucleate 
a bubble.  However, this statement involves an assumption about the 
spacetime measure on the multiverse, raising an unresolved issue in 
the understanding of eternal inflation.  We now briefly explain.

One consequence of eternal inflation is that the number of bubble 
nucleations (or of any type of event) diverges with time.  Indeed, 
the physical three-volume on an FRW foliation within a given bubble 
diverges as well.  Attempts to regulate these divergences have 
revealed that cosmological predictions tend to depend on the choice 
of regulator---this is known as the measure problem (for some reviews 
see e.g.~\cite{W06,G07}).  While it is unclear what is the 
correct measure on the multiverse, certain measures can be ruled out 
for making predictions in wild disagreement with our observations.  
In particular, measures that grant greater weight according to larger 
inflationary expansion factors suffer from the ``$Q$-catastrophe'' 
(and ``$G$-catastrophe'') \cite{FHW,GV05,GS}, predicting the 
amplitude of primordial density perturbations and the gravitational 
constant to be in stark conflict with what is observed.  Thus, one 
should be skeptical of arguments about the frequency of events in the 
multiverse based only on naive comparisons of inflationary expansion
factors.

In fact, the question of how the spacetime measure should address 
transdimensional tunneling in the multiverse has not yet been addressed 
in the literature.  To support the attitude expressed above, we here 
briefly speculate about three of the leading measure proposals.  One 
of these is the comoving probability measure \cite{Starobinsky,L06}, 
which can be seen as weighting events according to the frequency at 
which they are encountered by the future histories of a given 
worldline.  A simple generalization is to specify the worldline 
including its position in any initial compact dimensions, in which 
case this measure would not grant any additional weight to parent 
vacua simply because they have a greater number of expanding 
dimensions.  The generalization of the causal patch measure 
\cite{B06} is less clear, but since it also counts events according 
to their proximity to the future histories of a given worldline 
(specifically, whether or not they reside in the surrounding causal 
patch), one might guess that it would not give a very different result.  
The scale-factor cutoff measure \cite{L06,DSGSV} counts events only if 
they reside in the finite volume between an initial spacelike 
hypersurface and a later hypersurface determined by a fixed amount of 
expansion.  One way to extend this measure to the case of transdimensional
tunneling is to track the density of a fiducial ``dust'' of test 
particles scattered over the initial hypersurface, including over any 
compact dimensions, defining the later hypersurface according to when 
the density drops below a pre-specified value.  In this case, the 
extra volume due to an additional expanding dimension would be 
canceled by the extra dilution of the dust due to that expansion, and 
again parent vacua receive no additional weight simply due to having 
a greater number of expanding spatial dimensions.   

It is not enough to have an effectively (2+1)-dimensional parent 
vacuum nucleate a bubble of our vacuum phase---we also require an
appropriate period of slow-roll inflation subsequent to bubble
nucleation.  The analysis of this paper introduces a number of 
assumptions about this period of inflation, but we consider all but 
one to be simple matters of convenience.  For example, we assumed 
that the inflaton energy density is precisely constant during 
inflation, and that density perturbations are sourced by fluctuations 
in a second, subdominant field.  Yet we expect these choices to affect 
only the amplitude and tilt of the resulting spectrum of perturbations, 
both of which are model-dependent parameters that can presumably be 
tuned to match observation by simply picking an appropriate 
implementation of inflation.  

We also assumed that the inflaton energy density dominates over 
the modulus effective potential immediately after bubble nucleation.  
This was essential to obtaining a simple analytic solution for the 
background metric during inflation, but with hindsight we can see this 
too is not an important assumption.  The balance of contributions to
the energy density in the bubble affects only the time dependence of 
the scale factors $a$ and $b$, which in turn affects only the power 
spectrum $P_{qs}$ (not the anisotropic ``Fourier'' mode functions 
$U_{qrs}$).  The effect of anisotropy in the power spectrum was found 
to be subdominant to that coming from the mode functions; indeed the 
anisotropy of the power spectrum is ignored in the results of Section 
\ref{ssec:CMB}, since this allowed us to introduce an additional 
approximation to speed the numerical integration given our limited 
computational resources.       

The crucial assumption is that slow-roll inflation lasts long
enough to conform to present observation, but not so long as to 
push all of the effects of initial anisotropy outside of our present 
horizon.  This requires a sort of ``fine-tuning'' between the 
inflaton potential and the present age of the universe, which, at 
first glance, seems to involve an unusual coincidence.  However, such 
a ``coincidence'' may find an explanation in landscape cosmology. 

If indeed our pocket is one among a diverging set in an eternally 
expanding multiverse, then we must be careful to account for all of
the selection effects that modulate the likelihood for us to make a 
given observation.  These selection effects are ultimately 
encapsulated in the choice of spacetime measure; however in the 
present case it turns out that all three measures mentioned above 
give similar predictions \cite{FKRMS,BAP,BL09,omega}.  Roughly speaking,
there is a factor---called the ``prior''---which gives the relative 
probability that a random spacetime point resides in a bubble that 
undergoes $N$ $e$-folds of slow-roll inflation.  This is convoluted
with a second factor---called the ``anthropic factor''---which 
is proportional to the density of (appropriately defined) observers 
in bubbles characterized by $N$ $e$-folds of inflation.

The prior distribution is addressed in \cite{FKRMS}, where an 
(admittedly crude) argument is given to suggest that the landscape
might strongly prefer fewer $e$-folds of slow-roll inflation, with
the distribution $dP(N)\propto N^{-4}\,dN$.  The effect of anthropic 
selection is most carefully computed in \cite{omega}, where it is 
found that the mass function of Milky-Way mass galaxies is 
approximately independent of $\Omega_{\rm curv}$ for 
$\Omega_{\rm curv}\ll{\cal O}(0.1)$, and falls off exponentially for 
roughly $\Omega_{\rm curv}>0.7$ (note that 
$\Omega_{\rm curv}\propto e^{-2N}$ with a proportionality constant 
that depends on the scale of inflation and some details of reheating
and the subsequent big bang evolution).  The combination of these 
effects places most observers in spacetime regions in which 
$\Omega^0_{\rm curv}$ is unobservably small (i.e. below cosmic 
variance), but at the same time gives a reasonable probability, 
roughly of order ten percent chance, to observe 
$\Omega^0_{\rm curv}\gtrsim 10^{-5}$ \cite{BL09,omega}.

\section{Discussion}
\label{sec:discussion}

The present understanding of string theory and inflationary cosmology
points to a picture of spacetime containing countless bubbles endlessly
nucleating within an eternally-inflating background.  In the 
context of the string landscape, the complete set of bubbles contains 
a wide range of low-energy physics, including different numbers of 
compact dimensions (in fact it is the enormous variety of 
compactifications that inspires the diversity of the landscape in the 
first place).  It is therefore a priori possible that our bubble, 
containing three large (expanding) spatial dimensions, nucleated within 
a vacuum containing only two such dimensions.

For instance, the eternally-inflating vacuum in which our bubble 
nucleates could contain a compact dimension, the size of which is 
governed by a metastable modulus that tunnels through a potential 
barrier upon bubble nucleation.  The tunneling instanton and initial
bubble geometry then respect reduced symmetry from the (3+1)-dimensional
perspective, due to the presence of the additional compact dimension.  
Yet, as expected, a round of slow-roll inflation within the bubble is 
sufficient to redshift away the initial anisotropy and curvature, 
creating an O(3)-symmetric FRW cosmology consistent with the observed 
universe.  

Nevertheless, if inflation within the bubble does not last too long, 
effects of the initial anisotropy may be observable.  We here focus
on one such effect:  the generation of statistical anisotropy among 
large-scale inflationary perturbations.  We find that, when projected
onto a two-sphere approximating the surface of last scattering, the 
inflationary spectrum generates a multipole correlator 
$C_{\ell\ell'mm'}$ that features (in an appropriate coordinate system) 
off-diagonal elements in $\ell$ and $\ell'$ (when $\ell-\ell'=\pm 2$),
as well as dependence on the multipole moment $m$ (it is still diagonal
in $m$ and $m'$).  These effects are suppressed relative to the 
statistically-isotropic components of $C_{\ell\ell'mm'}$ by the 
present-day curvature parameter $\Omega^0_{\rm curv}$, but appear to 
extend to arbitrarily large $\ell$.  

There are a number of remaining issues to be explored.  Most 
importantly, as a first approach to the problem we have ignored the 
effects of spatial curvature and expansion anisotropy on the free 
streaming of photons from the surface of last scattering to the point
of present detection.  In fact, anisotropic spatial curvature sources
anisotropic expansion, which in turn deforms the surface of last
scattering away from the surface on which we project the inflationary
spectrum, in addition to perturbing the trajectories of geodesics as 
they radiate away from the point of observation.  A full understanding
of the observable signatures of anisotropic bubble nucleation requires 
combining both of these effects.

Also, we have ignored metric perturbations, focusing on the 
spectrum of a subdominant scalar field and assuming its isocurvature 
perturbations translate directly into adiabatic density perturbations.  
While we do not expect this to have a large effect on the spectrum of 
statistical anisotropies (compared to for instance the standard 
scenario where the primordial perturbations are sourced by the 
inflaton itself), it does not allow us to study the tensor 
perturbations generated during inflation.  Because scalar and tensor 
metric perturbations in general do not decouple in an anisotropic 
background, there are possibly interesting correlations between these 
signals.  Ignoring metric perturbations also hides any interesting 
effects that might come from fluctuations in the bubble wall itself.  

It would also be interesting to explore the nature and degree of  
non-Gaussianity implied by the existence of statistical anisotropy 
among inflationary perturbations.

Finally, another potential signature of multiverse cosmology is 
observable bubble collisions, see 
e.g.~\cite{JJBP,GGV,AJS,worldscollide,FKNS}.  
It would be interesting to understand whether the reduced symmetry at 
early times of anisotropic bubble nucleation affects the spectrum of 
bubble collisions on typical observer's sky, or if there is any 
special signature of collisions with bubbles containing a reduced 
number of large spatial dimensions \cite{MPS10}.

\acknowledgments

The authors thank Belen Barreiro, Raphael Bousso, Frederik Denef, Ben 
Freivogel, Jaume Garriga, Lam Hui, Matthew Johnson, Matthew Kleban, Ken 
Olum, Leonard Susskind, Vitaly Vanchurin, and Alexander Vilenkin for 
helpful discussions.  JJB-P and MPS are supported by the U.S. National 
Science Foundation, under grants 06533561 and 0855447, respectively.


%

\vspace{11pt}
\noindent
{\bf Note added:}  We very recently became aware of interesting work by
another group, which computes the effect of late-time spatial curvature 
and expansion anisotropy on photon free-streaming from the surface of
last scattering \cite{GHR}.

\appendix

\section{4d anisotropic tunneling instanton}
\label{sec:instanton}

We here verify the description of the tunneling instanton in the 
3d effective theory of Section \ref{ssec:instanton}, by solving for 
the corresponding solution in the full, anisotropic 4d geometry.  Our
starting point is the 4d action, (\ref{4daction}).  The 4d line 
element of the parent vacuum can be written
\beq
ds^2 = d\sigma^2 + a^2(\sigma)[-dt^2 + \cosh^2(t)\, d\phi^2] 
+ b^2(\sigma)\,dz^2 \,,
\label{appendixmetric}
\eeq
and as before we study the approximate complex-scalar-field solution 
$\varphi = \eta\,e^{inz}$.  Taking account of the energy momentum 
tensor of this model,
\beq
T_{\mu \nu} = K'\partial_{\mu} \varphi^*\partial_{\nu}\varphi
+ g_{\mu\nu} \left(-\frac{1}{2}K(X) - \frac{\lambda}{4}
\left(|\varphi |^2-\eta^2\right)^2 - \frac{\Lambda}{8\pi G}\right) ,
\eeq
where $X\equiv\partial_\mu\varphi^*\partial^\mu\varphi=n^2\eta^2b^{-2}$
and $K'\equiv dK/dX$, we obtain the equations of motion,
\bea
\frac{\dot{a}^2}{a^2} + 2\frac{\dot{a}\dot{b}}{a b} 
- \frac{1}{a^2} &=& -8 \pi G
\left(\frac{n^2 \eta^2}{2 b^2}+\frac{\kappa_2 n^4 \eta^4}{2
  b^4}+\frac{\kappa_3 n^6 \eta^6}{2 b^6}\right) - \Lambda \\
2\frac{\ddot{a}}{a} + \frac{\dot{a}^2}{a^2} - \frac{1}{a^2} &=&
8 \pi G \left(\frac{n^2 \eta^2}{2 b^2} + \frac{3 \kappa_2 n^4 \eta^4}
{2 b^4} + \frac{5\kappa_3 n^6 \eta^6}{2 b^6}\right)-\Lambda \,.
\eea
Above we have used $K(X)=X+\kappa_2X^2+\kappa_3X^3$ as in the main
text.  These equations of motion permit a solution of the form
\beq
a(\sigma) =  H_p^{-1} \sin(H_p\,\sigma) \,, \qquad 
b(\sigma) = b_p \,,
\eeq
with $H_p$ and $b_p$ being constants, provided that we impose the 
condition,
\beq
\frac{\Lambda}{8\pi G} + \frac{n^2 \eta^2}{b^2} 
+\frac{3\kappa_2 n^4 \eta^4}{2 b^4}
+\frac{4\kappa_3 n^6\eta^6}{2 b^6} = 0 \,,
\eeq
the solution of which is $b_d$.  The constant $H_p$ is then given by
\beq
H_p^2 = 8 \pi G
\left(\frac{n^2 \eta^2}{2 b_d^2}+\frac{\kappa_2 n^4 \eta^4}
{2b_d^4}+\frac{\kappa_3 n^6 \eta^6}{2 b_d^6}\right) + \Lambda \,.
\label{Hp}
\eeq
In the language of the 3d effective theory, this condition can be
written $\overline{V}' = 0$; therefore we have identified 
the compactification solution but from the vantage of the full 4d
theory.  As expected, inserting this solution into 
(\ref{appendixmetric}) gives the line element of (2+1)-dimensional 
de Sitter space crossed with a fixed-circumference circle.

\begin{figure}[t!]
\begin{center}
\begin{tabular}{cc}
\includegraphics[width=0.475\textwidth]{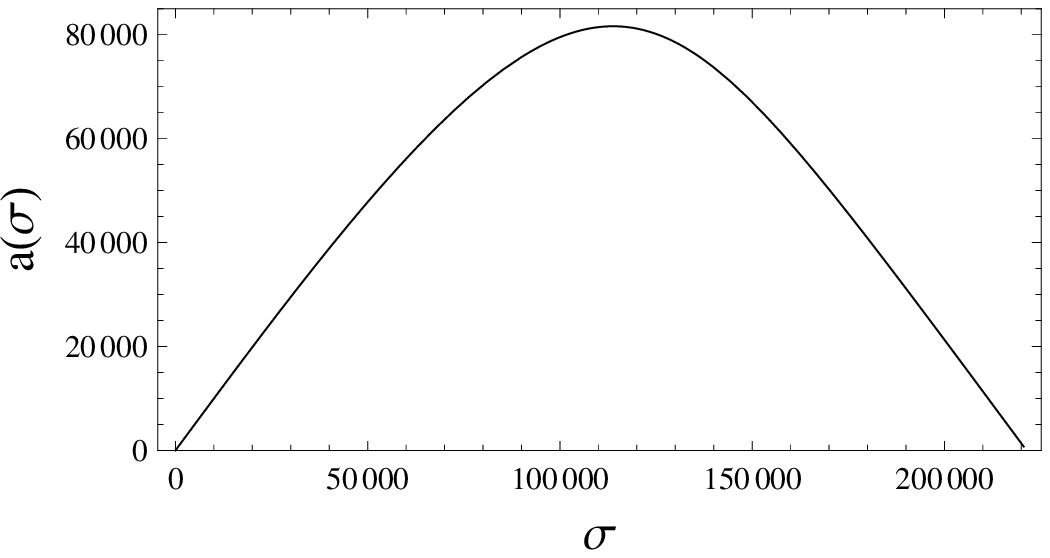} &
\includegraphics[width=0.457\textwidth]{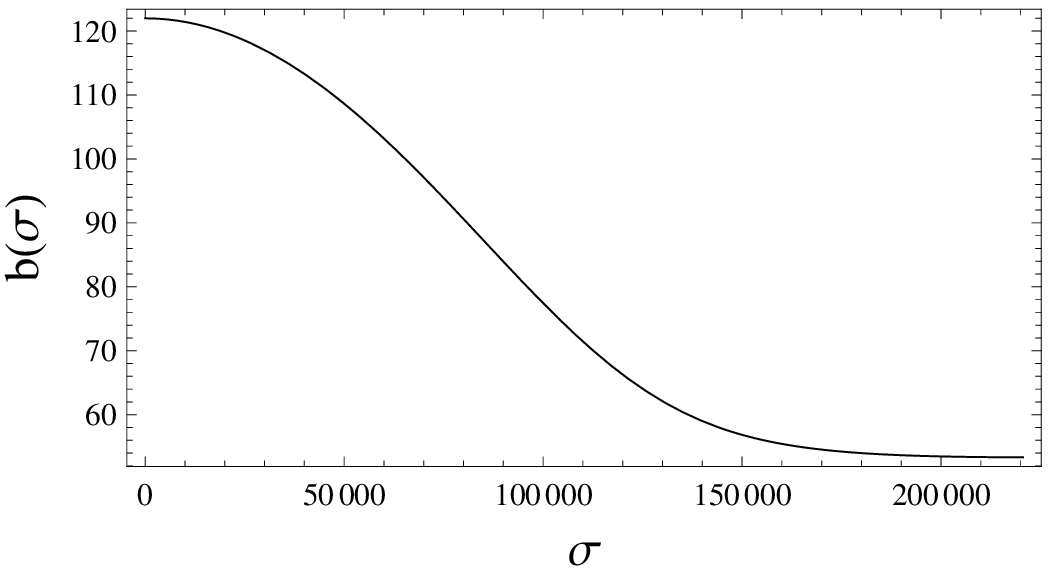} \\
\end{tabular}
\caption{\label{fig:instanton4d}The 4d instanton solution for $a(\sigma)$ 
(left panel) and $b(\sigma)$ (right panel), all quantities given in units 
of $G$.  The large numbers are due to the dynamics being 
sub-Planckian; see text for details.}
\end{center}
\end{figure}

Consider evolving the equations of motion numerically, taking the 
boundary conditions 
\bea
a(\sigma) &=& \sigma + 2\pi G\left(\frac{n^2\eta^2}{9b_d^2}
+\frac{\kappa_2n^4\eta^4}{3b_d^4}
+\frac{5\kappa_3n^6\eta^6}{9b_d^6}
-\frac{\Lambda}{36\pi G}\right) \sigma^3 + \ldots \\
b(\sigma) &=& b_d - 2\pi G\left(\frac{2n^2\eta^2}{3b_d}
+\frac{\kappa_2n^4\eta^4}{b_d^3}
+\frac{4\kappa_3n^6\eta^6}{3b_d^5}
-\frac{\Lambda}{12\pi G}\right)\sigma^2 + \ldots \,,
\eea
in the limit $\sigma \rightarrow 0$.  This corresponds to a Taylor
expansion of $a(\sigma)$ and $b(\sigma)$, taking $b(0)=b_d$, and 
obtaining the other coefficients of the expansion by inserting into
the equations of motion.  Given these boundary conditions, one can 
show the geometry is smooth in the limit $\sigma\rightarrow 0$.  We 
then set the value of $b_d$ by trial and error, so that the entire
solution is smooth, in particular so that as $\sigma$ approaches 
some value $\sigma_{\rm max}$, $a\to\sigma_{\rm max}-\sigma$ and 
$b\to$ constant.    
The results of such a numerical evolution are displayed in Figure
\ref{fig:instanton4d}.  To generate these curves, we have used the
same values of parameters as are used to generate Figures 
\ref{fig:V} and \ref{fig:instanton3d} in Section \ref{sec:bubble}.  
The results agree with those of Section \ref{ssec:instanton} in 
the sense that when we read off the initial and final values of 
the circumference of the $z$ dimension, between the two approaches
they agree.

While the interpretation of this instanton as a tunneling event is 
not so apparent in the 4d picture, note that the same instanton 
would describe the reverse ``tunneling'' process, the description of 
which is less clear in the 3d effective theory, which breaks down
inside the nucleating bubble.  (While we do not include the evolution
here, there is no difficulty in extending the evolution of $a$ and $b$
to convey the evolution in the daughter vacuum.)  Viewing this 
solution as a tunneling instanton, one can interpret it as an inflating 
black brane, charged with respect to the scalar field $\varphi$, 
nucleating in the 4d de Sitter background spacetime. Higher dimensional 
solutions analogous to this have been discussed by \cite{B-PS-PV-2}.

\section{Bunch--Davies vacuum}
\label{sec:BD}

\begin{figure}[t!]
\begin{center}
\includegraphics[width=0.3\textwidth]{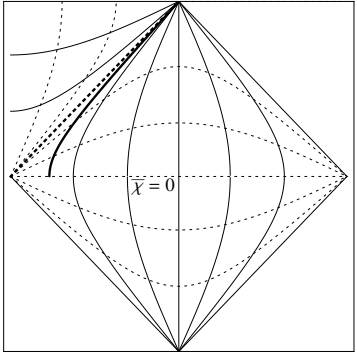} 
\caption{\label{fig:conformal}Conformal diagram of a de Sitter 
bubble nucleating in a de Sitter parent vacuum, indicating the 
bubble wall (thick solid curve), the future lightcone of the bubble 
nucleation event (thick dashed curve), some surfaces of constant 
$\eta_c$ and $\be$ (solid curves), and some surfaces of constant 
$\chi$ and $\bc$ (dashed curves).  The uncharted regions of spacetime 
are not relevant to our discussion.  The Cauchy surface $\bc=0$ is 
indicated.}
\end{center}
\end{figure}

To determine the power spectrum of the light scalar $\sigma$ with 
respect to the Bunch--Davies vacuum \cite{BunchDavies},  
we trace the evolution of $\Upsilon^\upsilon_{qs}$ back into the parent 
vacuum and Klein-Gordon normalize the positive-frequency modes over a 
Cauchy surface, as is done in \cite{BGT,YST,GMST}.  Figure 
\ref{fig:conformal} displays a conformal diagram illustrating the 
geometry (the coordinates $z$ and $\phi$ are suppressed).  For 
simplicity we have constructed the diagram as if the vacuum energies 
inside and outside of the bubble are the same; however the statements 
below apply equally to more realistic geometries.  We have also drawn 
the diagram (and we work below) as if all of spacetime can be covered 
by just these two universes---in other words, as if there were no other 
bubbles.\footnote{In fact the daughter bubble experiences an 
infinite number of collisions with other bubbles that nucleate in the
parent vacuum \cite{GGV,AJS,worldscollide,FKNS}; however the 
phenomenological success of inflationary theory suggests that any 
effects of such collisions should be small---linear perturbations to 
the standard inflationary background (at least in our past lightcone).  
Meanwhile, the parent vacuum itself likely resides in an open bubble, 
or contains other bubbles that intersect the Cauchy surface at $\bc=0$ 
(see text below).  However bubble nucleation rates are likely to be 
exponentially suppressed, so any such disturbances would typically be 
very far from the physical region of interest.  For simplicity we here 
presume both effects are negligible.}  
Within the bubble, which nucleates at the middle of the left boundary 
and thus occupies the upper left corner of the diagram, $\chi$ is a 
radial coordinate on constant-$z$ (2+1)-dimensional open FRW 
hypersurfaces, and $\eta_c$ provides a spacelike foliation.  Evidently 
no Cauchy surface can be drawn within the bubble.  

To simplify the geometry of the parent vacuum we assume the energy 
density of the bubble wall does not significantly perturb the metric.  
Furthermore, we treat the radius $b_d$ of the compact dimension $z$ 
as if it were essentially constant throughout the parent vacuum, and 
we take the energy density immediately inside the bubble to be 
essentially equal to that of the parent vacuum on the outside (to be
precise, we take $H_{\rm inf}^2=\Lambda/3$ to be essentially the
same as the $H_p^2$ of (\ref{Hp})).  Note that these are not 
necessarily good approximations of the toy compactification 
model described in Section \ref{sec:bubble}; however they greatly 
simplify the computations here.  In particular they allow us to treat 
the parent vacuum as pure de Sitter space (crossed with a circle), in 
which case the central, diamond-shaped region of Figure 
\ref{fig:conformal} can be covered by coordinates with line element
\beq
ds^2 = \ba^2(\be)\left[d\be^2-d\bc^2+\cosh^2(\bc)\,d\phi^2\right]
+b_d^2\,dz^2 \,.
\label{ametric}
\eeq
Here the coordinate $\bc$ provides the spacelike foliation, $\be$ is 
a spatial coordinate, and the hypersurface $\bc=0$ is a Cauchy surface.  
Since $\bc=0$ lies entirely within the region covered by 
(\ref{ametric}), it is unnecessary to cover any more of the spacetime 
by introducing additional charts. 

Note that the bubble coordinate system of (\ref{cmetric}) can be 
smoothly connected to the parent vacuum coordinate system of 
(\ref{ametric}) by using analytic continuation, 
\beq
\bc = \chi - \frac{i\pi}{2} \,, \qquad
\be = \eta_c + \frac{i\pi}{2} \,, \qquad
\ba=ia \,.
\label{ac}
\eeq 
Here the use of analytic continuation is merely a technical device 
to describe the propagation of modes through the bubble wall.  In
particular, we do not transform the coordinate of the bubble scale
factor $b(\eta_c)$.  Instead the $z$ dimension of the parent vacuum maps 
directly onto that of the bubble (as is the case with the coordinate
$\phi$), assuming that $b(\eta_c)\to b_d$ and $\dot{b}_d\to 0$ as 
$\eta_c\to-\infty$.

Within the parent vacuum we denote the mode expansion of $\sigma$ 
according to 
\beq
\sigma_{\upsilon qrs}(\be,\,\bc,\,\phi,\,z) = 
\frac{\cN_q}{\sqrt{\ba\,b_d}}\,\bU^\upsilon_{qs}(\be)\,
\overline{U}_{qrs}(\bbfx)
=\frac{\cN_q}{\sqrt{\ba\,b_d}}\,\bU^\upsilon_{qs}(\be)\,
\overline{X}_{qr}(\bc)\,\overline{\Phi}_r(\phi)\,\overline{Z}_s(z)\,,
\label{cmodes}
\eeq
where to simplify the presentation here and below we do not worry 
about keeping track of the (unimportant) overall phase.  We have 
introduced the superfluous normalization factor $\cN_q$ for later 
convenience.  The $\phi$- and $z$-dependent Fourier modes are the 
same as those within the bubble, 
$\overline{\Phi}_r(\phi) = \Phi_r(\phi)$ and 
$\overline{Z}_s(z) = Z_s(z)$.  Meanwhile, the $\bc$-dependent modes 
can still be written as solutions to the Legendre equation, but they 
now take the form  
\beq
\overline{X}_{qr}(\bc) = 
\frac{\Gamma(\frac{1}{2}+iq-r)}{\Gamma(iq)}\,
P^r_{iq-1/2}(i\sinh(\bc)) \,.
\label{Xsol2}
\eeq
The normalizations of $\overline{\Phi}_r$, $\overline{Z}_s$, and
$\overline{X}_{qr}$ are such that, after analytic continuation, 
each mode maps onto the similarly-denoted mode defined within the 
bubble.  Note that the modes $\overline{X}_{qr}$, which describe 
the time-like evolution in the parent vacuum, correspond to 
positive-frequency plane waves $\propto e^{-iq\bc}$ (up to an 
unimportant phase) in the limit $\bc\to 0$.  

The analytic continuation of the scale factor $a$ gives 
$\ba=H_{\rm inf}^{-1}\,{\rm sech}(\be)$, while in the parent 
vacuum the scale factor $b$ is constant, $b=b_d$.  This gives the 
equation of motion
\beq
\ddot{\bU}^\upsilon_{qs} +\left[q^2+\left(\frac{3}{4}-\mu^2\right)
{\rm sech}^2(\be)\right] \bU^\upsilon_{qs} = 0\,,
\label{SE}
\eeq
where the dots now denote derivatives with respect to $\be$, and
again $\mu=s/b_dH_{\rm inf}$.  Notice that the separation constant 
$\mu^2$ appears as would a mass for the scalar field $\sigma$.  The 
general solution of (\ref{SE}) can be written in terms of associated 
Legendre functions:
\bea
\bU^\upsilon_{qs} &=& \bC^\upsilon_1\, 
P^{iq}_{\sqrt{1-\mu^2}-1/2}(\tanh(\be))
+ \bC^\upsilon_2\, P^{-iq}_{\sqrt{1-\mu^2}-1/2}(\tanh(\be)) 
\label{sols2} \\
&\equiv& \bC^\upsilon_1\,\bcF_1(\be)+\bC^\upsilon_2\,\bcF_2(\be) \,. 
\phantom{P_{i}^{\sqrt{1-\mu^2}}} \label{bcFdef}
\eea

The solutions $\bcF_1$ and $\bcF_2$ represent two orthogonal spatial 
modes, and thus each should be normalized separately, with the two  
summed over in the final mode expansion.  Hence we have introduced 
the index $\upsilon$, which distinguishes these modes.  To perform 
the normalization it is convenient to apply a trick.  Notice that 
(\ref{SE}) has the form of a Schr\"odinger equation for a point 
particle with position $\be$, potential energy 
$V(\be)=(\mu^2-3/4)\,{\rm sech}^2(\be)$, and energy $q^2$.  The 
potential $V(\be)$ tends to zero as $\be\to\pm\infty$; thus 
$\bU^\upsilon_{qs}$ may just as well be expressed in terms of two 
linearly independent ``scattering'' solutions of the form
\bea
i\bU_{qs}^- &\to& \left\{ 
\begin{array}{ll}
\omega_-\, e^{-iq\be} + e^{iq\be} \qquad  
& {\rm as}\quad \be\to -\infty \\
\varsigma_-\, e^{iq\be} & {\rm as}\quad \be\to +\infty \,,
\end{array}
\right. \label{nsol} \\
i\bU_{qs}^+ &\to& \left\{ 
\begin{array}{ll}
\varsigma_+\, e^{-iq\be} & {\rm as}\quad \be\to -\infty\,,\\
\omega_+\, e^{iq\be} + e^{-iq\be}\qquad  
& {\rm as}\quad \be\to +\infty \,,
\end{array}
\right. \label{psol} 
\eea
where studying the Wronskian reveals 
$|\omega_\pm|^2+|\varsigma_\pm|^2=1$,\, $\varsigma_+=\varsigma_-$,
and $\varsigma_+\omega_-^*+\varsigma_-^*\omega_+=0$ \cite{GMST}.  (The
factor $\cN_q$ was introduced in (\ref{bubmodedef}) and 
(\ref{cmodes}) to permit these simple normalizations; it will be 
determined later.)  The normalization of the delta function is 
determined by noting that all of its support comes from large $|\be|$, 
where the solutions are plane-wave.  Thus 
\beq
\int d\be \bU_{qs}^\upsilon\,\bU_{q's}^{\upsilon' *} = 2\pi\,\delta(q-q')\,
\delta_{\upsilon\upsilon'} \,,
\label{Unorm}
\eeq  
where the index $\upsilon$ takes values $\pm$.  We have introduced 
the modes $\bU_{qs}^\pm$ merely to take advantage of this
simple normalization.    

The normalized scattering solutions $\bU_{qs}^\pm$ can be expressed 
in terms of the solutions $\bcF_1$ and $\bcF_2$ by matching their
asymptotic behavior onto (\ref{nsol}) and (\ref{psol}).  In 
particular, we write
\bea
\begin{array}{ll}
\bcF_1(\be) \,\to\, \alpha^-_1\,e^{-iq\be}+\beta^-_1\,e^{iq\be} \quad
\phantom{\displaystyle\Big(\Big)} & {\rm as} \quad \be\to -\infty \\
\bcF_1(\be) \,\to\, \alpha^+_1\,e^{-iq\be}+\beta^+_1\,e^{iq\be} \quad
\phantom{\displaystyle\Big(\Big)} & {\rm as} \quad \be\to +\infty \\
\bcF_2(\be) \,\to\, \alpha^-_2\,e^{-iq\be}+\beta^-_2\,e^{iq\be} \quad
\phantom{\displaystyle\Big(\Big)} & {\rm as} \quad \be\to -\infty \\
\bcF_2(\be) \,\to\, \alpha^+_2\,e^{-iq\be}+\beta^+_2\,e^{iq\be} \quad
\phantom{\displaystyle\Big(\Big)} & {\rm as} \quad \be\to +\infty\,, \\
\end{array}
\label{bFlim}
\eea  
where the coefficients $\alpha_i^\pm$ and $\beta_i^\pm$ are given by
\bea
\alpha^-_1 &=& 
\frac{e^{-\pi q/2}\,\Gamma(iq)}
{\Gamma(\frac{1}{2}+\sqrt{1-\mu^2})\,
\Gamma(\frac{1}{2}-\sqrt{1-\mu^2})} \label{ba1}\\
\beta^-_1 &=& 
\frac{e^{-\pi q/2}\,\Gamma(-iq)}
{\Gamma(\frac{1}{2}-iq+\sqrt{1-\mu^2})\,
\Gamma(\frac{1}{2}-iq-\sqrt{1-\mu^2})} \\
\alpha^-_2 &=& 
\frac{e^{\pi q/2}\,\Gamma(iq)}
{\Gamma(\frac{1}{2}+iq+\sqrt{1-\mu^2})\,
\Gamma(\frac{1}{2}+iq-\sqrt{1-\mu^2})} \\
\beta^-_2 &=& 
\frac{e^{\pi q/2}\,\Gamma(-iq)}
{\Gamma(\frac{1}{2}+\sqrt{1-\mu^2})\,
\Gamma(\frac{1}{2}-\sqrt{1-\mu^2})} \\
\alpha^+_1 &=& 0 \,,\quad
\beta^+_1 = \frac{e^{-\pi q/2}}{\Gamma(1-iq)}\,,\quad
\alpha^+_2 = \frac{e^{\pi q/2}}{\Gamma(1+iq)}\,,\quad
\beta^+_2 =  0 \label{ba2}\,.
\eea
By matching the asymptotic positive and negative ``frequency'' modes 
of $\bcF_1$ and $\bcF_2$ onto the asymptotic behavior of 
$\bU_{qs}^+$ and $\bU_{qs}^-$ given in (\ref{nsol}) and (\ref{psol}),
we determine
\bea
\bU_{qs}^-(\be) &=& \frac{-i\alpha^+_2\,\bcF_1(\be)+i\alpha^+_1\,\bcF_2(\be)}
{\alpha^+_1\,\beta^-_2-\alpha^+_2\,\beta^-_1}\\
\bU_{qs}^+(\be) &=& \frac{i\beta^-_2\,\bcF_1(\be)-i\beta^-_1\bcF_2(\be)}
{\alpha^+_1\,\beta^-_2-\alpha^+_2\,\beta^-_1} \,.
\eea
(Some of the terms above are zero, however we delay simplification
until later.)

We have now determined all of the mode functions, up to the overall
normalization $\cN_q$.  This is determined by enforcing Klein-Gordon 
normalization of the scalar field $\sigma$, 
\bea
(\sigma_{\upsilon qrs},\,\sigma_{\upsilon'q'r's'}) 
&\equiv& -i\int d\Sigma_\mu\, g^{\mu\nu}
\left[ \sigma_{\upsilon qrs}\,\partial_\nu\sigma^*_{\upsilon'q'r's'}-
\left(\partial_\nu\sigma_{\upsilon qrs}\right)
\sigma^*_{\upsilon'q'r's'}\right] \nn\\
&=& \,\delta(q-q')\,\delta_{rr'}\,\delta_{ss'}\,
\delta_{\upsilon\upsilon'} \,.
\label{KGnorm}
\eea
As mentioned before, we set the Cauchy hypersurface $\Sigma_\mu$ at 
$\bc=0$.  Using the orthogonality of the mode functions $\bU_{qs}^\pm$, 
$\overline{\Phi}_r$, and $\overline{Z}_s$, the above normalization 
condition can be written
\beq
2\pi\,|\cN_q|^2\cosh(\bc)\left[\overline{X}_{qr}\,
\partial_{\bc}\overline{X}_{qr}^*-\left(\partial_{\bc}
\overline{X}_{qr}\right)\overline{X}_{qr}^*\right]
\!\Big|_{\bc=0} 
= i \,.
\eeq
The term in brackets is at first glance rather complicated; however it
can be simplified with some technical manipulations, and in the end we 
find (up to an unimportant phase)
\beq
\cN_q  = \frac{1}{\sqrt{4q\sinh(\pi q)}}\,. 
\eeq

Now that we have the Bunch--Davies modes of $\sigma$ in the parent
vacuum, the next step is to propagate these modes into the bubble.
For the modes $\overline{\Phi}_r$ and $\overline{Z}_s$ this is easy: 
they are unchanged.  The modes $\overline{X}_{qr}$ and $\bU^\pm_{qs}$ 
are propagated by analytic continuation of the coordinates in their 
arguments, as given by (\ref{ac}).  The temporal mode 
functions $\overline{X}_{qr}(\bc)$ become the bubble spatial modes 
$X_{qr}(\chi)$ of (\ref{Xmodes}).  It is left to discuss the 
modes $\bU^\pm_{qs}$.

Note that the only dependence of the $\bU_{qs}^\pm$ on $\be$ is via 
$\tanh(\be)$, which is rotated into $\coth(\eta_c)$ when we take 
$\be\to\eta_c+i\pi/2$.  The functions $\bcF_1$ and $\bcF_2$ thus
become
\bea
\bcF_1(\be) \,&\to&\, \btF_1(\eta_c) = 
P^{iq}_{\sqrt{1-\mu^2}-1/2}(\coth(\eta_c))\\
\bcF_2(\be) \,&\to&\, \btF_2(\eta_c) =
P^{-iq}_{\sqrt{1-\mu^2}-1/2}(\coth(\eta_c)) \,.
\eea  
These are not the same as the functions $\cF_1$ and $\cF_2$ 
computed with the bubble line element (\ref{cmetric}), because 
those solutions account for the growth in the scale factor 
$b(\eta_c)$, whereas $\btF_1$ and $\btF_2$ continue from the parent 
vacuum where $b$ is static.  Nevertheless they have the correct 
asymptotic form, as $\eta_c\to-\infty$, because the scale factor 
$b(\eta_c)\propto\coth(\eta_c)$ approaches a constant in that limit.  
All we must do is match the solutions $\cF_1$ and $\cF_2$ onto $\btF_1$ 
and $\btF_2$, in the limit $\eta_c\to-\infty$.  In this limit, $\btF_1$ 
and $\btF_2$ have the asymptotic behavior
\beq
\btF_1(\eta_c)\,\to\, \tilde{\alpha}_1\,e^{-iq\eta_c} 
+ \tilde{\beta}_1\,e^{iq\eta_c} \,, \qquad
\btF_2(\eta_c)\,\to\, \tilde{\alpha}_2\,e^{-iq\eta_c} 
+ \tilde{\beta}_2\,e^{iq\eta_c} \,,
\label{tmodes}
\eeq   
where the coefficients $\tilde{\alpha}_i$ and $\tilde{\beta}_i$ are
given by
\bea
\tilde{\alpha}_1 &=& 
e^{\pi q/2}\alpha^-_1 = 
\frac{\Gamma(iq)}{\Gamma(\frac{1}{2}+\sqrt{1-\mu^2})\,
\Gamma(\frac{1}{2}-\sqrt{1-\mu^2})} \label{ta1}\\
\tilde{\beta}_1 &=&  
e^{-\pi q/2}\beta^-_1 = 
\frac{e^{-\pi q}\,\Gamma(-iq)}{\Gamma(\frac{1}{2}-iq+\sqrt{1-\mu^2})\,
\Gamma(\frac{1}{2}-iq-\sqrt{1-\mu^2})} \label{tb1}\\ 
\tilde{\alpha}_2 &=& 
e^{\pi q/2}\alpha^-_2 = 
\frac{e^{\pi q}\,\Gamma(iq)}{\Gamma(\frac{1}{2}+iq+\sqrt{1-\mu^2})\,
\Gamma(\frac{1}{2}+iq-\sqrt{1-\mu^2})} \label{ta2}\\
\tilde{\beta}_2 &=& 
e^{-\pi q/2}\beta^-_2 = \frac{\Gamma(-iq)}
{\Gamma(\frac{1}{2}+\sqrt{1-\mu^2})\,
\Gamma(\frac{1}{2}-\sqrt{1-\mu^2})} \label{tb2}\,.
\eea

The functions $\cF_1$ and $\cF_2$ have been expressed so as to have
simple asymptotic behavior, see (\ref{F12lim1}).  Putting all of this
together, we obtain the pair of mode functions 
\bea
\Upsilon_{qs}^-(\eta_c) &=& 
\left(\frac{\alpha^+_1\tilde{\alpha}_2-\alpha^+_2\tilde{\alpha}_1}
{\alpha^+_1\,\beta^-_2-\alpha^+_2\,\beta^-_1}\right)i\cF_1(\eta_c)
+\left(\frac{\alpha^+_1\tilde{\beta}_2-\alpha^+_2\tilde{\beta}_1}
{\alpha^+_1\,\beta^-_2-\alpha^+_2\,\beta^-_1}\right)i\cF_2(\eta_c)
\label{ps1}\\
\Upsilon_{qs}^+(\eta_c) &=& 
\left(\frac{\beta^-_2\tilde{\alpha}_1-\beta^-_1\tilde{\alpha}_2}
{\alpha^+_1\,\beta^-_2-\alpha^+_2\,\beta^-_1}\right)i\cF_1(\eta_c)
+\left(\frac{\beta^-_2\tilde{\beta}_1-\beta^-_1\tilde{\beta}_2}
{\alpha^+_1\,\beta^-_2-\alpha^+_2\,\beta^-_1}\right)i\cF_2(\eta_c)
\label{ps2}\,.
\eea
Referring to (\ref{ta1})--(\ref{tb2}), we see that the terms in 
the first parantheses of (\ref{ps1}) cancel to zero, as do the terms
in the second parantheses of (\ref{ps2}).  Referring also to 
(\ref{ba1})--(\ref{ba2}), we see that other basic simplifications are
possible, and in the end we obtain
\bea
\Upsilon_{qs}^-(\eta_c) &=& ie^{-\pi q/2} \cF_2(\eta_c) \\
\Upsilon_{qs}^+(\eta_c) &=& ie^{\pi q/2}
\left(\frac{\beta^-_1\alpha^-_2-\beta^-_2\alpha^-_1}
{\alpha^+_2\,\beta^-_1}\right)\cF_1(\eta_c) \,.
\eea

The power spectrum can now be computed just as in Section 
\ref{ssec:power}, except now we must include the extra normalization 
factor $\cN_q$.  That is, we write
\beq
\<\hat{\sigma}(\eta_c,\,q,\,r,\,s)\,
\hat{\sigma}^\dagger(\eta_c,\,\q,\,\r,\,\s)\>
= \frac{\cN_q^2}{a(\eta_c)b(\eta_c)}\sum_\upsilon
|\Upsilon_{qs}^\upsilon(\eta_c)|^2\,
\delta(q-\q)\,\delta_{r\r}\,\delta_{s\s} \,. 
\label{power2}
\eeq
As described in Section \ref{ssec:power} we evaluate the power spectrum 
in the limit $\eta_c\to 0$, in which case the asymptotic behavior of 
the time-dependent parts is given by (\ref{F1lim2}) and 
(\ref{F2lim2}).  Putting everything together, and performing some 
manipulations to eliminate some of the gamma functions, we find the 
power spectrum to be
\bea
P_{qs} &=& \lim_{\eta_c\to 0}\, 
\frac{\cN_q^2}{a(\eta_c)b(\eta_c)}
\sum_{\upsilon}|\Upsilon_{qs}^\upsilon(\eta_c)|^2 \\
&=& \frac{\pi H_{\rm inf}}{8\,b_d\sinh^2(\pi q)}
\left\{\pi\cosh(\pi q)
\left|\Gamma\!\left(\frac{5}{4}+\frac{iq}{2}+\frac{\mu}{2}\right)\right|^{-2}\,
\left|\Gamma\!\left(\frac{5}{4}+\frac{iq}{2}-\frac{\mu}{2}\right)\right|^{-2}
\right.\nn\\
&& -\left. \cos(\pi\sqrt{1-\mu^2})
{\rm Re}\!\left[\frac{2^{2iq}\,\Gamma(\frac{1}{2}-iq-\sqrt{1-\mu^2})\,
\Gamma(\frac{1}{2}-iq+\sqrt{1-\mu^2})}
{\Gamma^2(\frac{5}{4}-\frac{iq}{2}-\frac{\mu}{2})\,
\Gamma^2(\frac{5}{4}-\frac{iq}{2}+\frac{\mu}{2})}\right]\right\} . \qquad
\label{truepower}
\eea

\begin{figure}[t!]
\begin{center}
\includegraphics[width=0.5\textwidth]{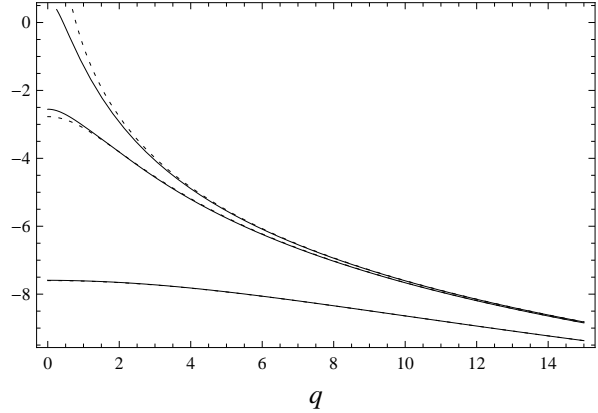}
\caption{\label{fig:spectrum}The power spectrum $P_{qs}$, 
expressed as $\ln(b_dP_{qs}/H_{\rm inf})$, plotted as a function 
of $q$ for $\mu=0.1$ (top pair of curves), $\mu=2$ (middle pair of 
curves), and $\mu=10$ (bottom pair of curves).  In each case we plot
the actual power spectrum (solid) and the ``asymptotic limit'' 
referred to in the text (dashed).}
\end{center}
\end{figure}

In Figure \ref{fig:spectrum} we plot the above power spectrum, and its 
small scale (large $q$, $\mu$) asymptotic limit (\ref{estimate}), for 
a few values of $\mu$.  It is clear that the spectrum rapidly approaches 
its asymptotic limit, with significant deviations only for 
$q\lesssim {\cal O}(1)$.  This is not unlike the situation in regular 
(isotropic) 4d open bubble inflation, where the inflationary spectrum 
is also modified at $q\lesssim {\cal O}(1)$ \cite{BGT,YST,GMST}.  In 
the present situation, it is evident that the deviation from the 
asymptotic curve is itself anisotropic; for example at $q=2$ the curve 
with $\mu=0.1$ deviates more significantly than the curve with 
$\mu=10$.  This anisotropy is on top of that which results from the
use of anisotropic mode functions in the power spectrum.

\end{document}